\documentclass[twocolumn,trackchanges]{aastex7}
\usepackage{amsmath}
\usepackage{subfigure}

\shortauthors{Ma et al.}

\begin{document}
\defcitealias{Kainulainen2009}{K09}
\defcitealias{Kolmogorov1941a}{K41}
\defcitealias{She1994}{SL94}
\defcitealias{Boldyrev2002a}{B02}
\title{Examining Turbulence in Galactic Molecular Clouds. II. Turbulence Cascade Beyond the Scale of Giant Molecular Clouds}

\correspondingauthor{Yuehui Ma, Xuepeng Chen, Hongchi Wang}

\author[orcid=0000-0002-8051-5228, sname='Ma']{Yuehui Ma}
\affiliation{Purple Mountain Observatory and Key Laboratory of Radio Astronomy, Chinese Academy of Sciences, 10 Yuanhua Road, Nanjing 210033, China}
\email[show]{mayh@pmo.ac.cn}  

\author[orcid=0000-0002-6388-649X, gname=Miaomiao, sname='Zhang']{Miaomiao Zhang} 
\affiliation{Purple Mountain Observatory and Key Laboratory of Radio Astronomy, Chinese Academy of Sciences, 10 Yuanhua Road, Nanjing 210033, China}
\email{miaomiao@pmo.ac.cn}

\author[orcid=0000-0003-0746-7968, gname=Hongchi, sname='Wang']{Hongchi Wang} 
\affiliation{Purple Mountain Observatory and Key Laboratory of Radio Astronomy, Chinese Academy of Sciences, 10 Yuanhua Road, Nanjing 210033, China}
\email[show]{hcwang@pmo.ac.cn}

\author[orcid=0000-0003-3151-8964, gname=Xuepeng, sname='Chen']{Xuepeng Chen} 
\affiliation{Purple Mountain Observatory and Key Laboratory of Radio Astronomy, Chinese Academy of Sciences, 10 Yuanhua Road, Nanjing 210033, China}
\affiliation{School of Astronomy and Space Science, University of Science and Technology of China, Hefei, Anhui 230026, China}
\email[show]{xpchen@pmo.ac.cn}

\author[orcid=0009-0009-3431-1150, gname=Zhenyi, sname='Yue']{Zhenyi Yue} 
\affiliation{Purple Mountain Observatory and Key Laboratory of Radio Astronomy, Chinese Academy of Sciences, 10 Yuanhua Road, Nanjing 210033, China}
\affiliation{School of Astronomy and Space Science, University of Science and Technology of China, Hefei, Anhui 230026, China}
\email{zyyue@pmo.ac.cn}

\author[orcid=0009-0004-2947-4020, gname=Suziye, sname='He']{Suziye He} 
\affiliation{Purple Mountain Observatory and Key Laboratory of Radio Astronomy, Chinese Academy of Sciences, 10 Yuanhua Road, Nanjing 210033, China}
\affiliation{School of Astronomy and Space Science, University of Science and Technology of China, Hefei, Anhui 230026, China}
\email{hszy@pmo.ac.cn}

\author[orcid=0009-0009-8707-8620, gname=Xiangyu, sname='Ou']{Xiangyu Ou} 
\affiliation{Purple Mountain Observatory and Key Laboratory of Radio Astronomy, Chinese Academy of Sciences, 10 Yuanhua Road, Nanjing 210033, China}
\affiliation{School of Astronomy and Space Science, University of Science and Technology of China, Hefei, Anhui 230026, China}
\email{oxy@pmo.ac.cn}

\author[orcid=0000-0003-2732-0592, gname=Li, sname='Sun']{Li Sun} 
\affiliation{Purple Mountain Observatory and Key Laboratory of Radio Astronomy, Chinese Academy of Sciences, 10 Yuanhua Road, Nanjing 210033, China}
\email{lisun@pmo.ac.cn}

\begin{abstract}
We use $^{12}$CO (J=1--0) data from the MWISP survey to investigate turbulence in a $\sim$kpc-scale segment of the Local Arm. By slicing the position-position-velocity cube into narrow layers that follow the Galactic-rotation trend in the $L$-$V$ diagram, we find that the structure functions (SFs) and spatial power spectra (SPS) of the $^{12}$CO (J=1--0) line velocity and intensity exhibit consistent scaling behaviors across all layers, demonstrating that the molecular gas forms a single, coherent turbulent field with energy cascading from $\sim 200$~pc down to parsec scales. The SPS power-law slopes of both the intensity and velocity fields approach the values expected from turbulence models. Cloud-to-cloud velocity SFs based on the molecular clouds located in the region follow the same extended self-similarity scaling (ESS) as the pixel-based statistics, indicating that inter-cloud motions are part of the same large-scale turbulent cascade. Together, these results provide direct observational evidence that molecular clouds are not dynamically isolated entities but are embedded within a larger-scale turbulent flow that links galactic dynamics to star formation on cloud scales.

\end{abstract}

\section{Introduction} \label{sec1}
Turbulence is a fundamental property of fluids, with kinetic energy injected on large scales and cascading toward small scales through a hierarchy of eddies \citep{Kolmogorov1941a, Frisch1995}. While terrestrial studies often focus on controlling or modeling turbulent flows \citep{Pope2000}, studies of astrophysical turbulence typically center on characterizing the basic properties of turbulence through observations. In the interstellar medium (ISM), turbulence is inherently multiscale and multiphase \citep{Kobayashi2022}, and becomes supersonic in the molecular clouds where stars form. It plays an important role in shaping the molecular clouds and in the regulation of star formation \citep{ElmegreenScalo2004, MacLowKlessen2004}. However, there still remains active debate over whether turbulence, gravity, or magnetic fields dominate molecular cloud dynamics and star formation \citep{Vazquez-Semadeni2025}. Thus, observationally determining key turbulent properties$-$its type, driving mechanisms, dissipation scale, and intermittency$-$is essential for clarifying its role in star formation.

The most fundamental statistical characteristic of turbulence is its power-law energy spectrum within the so-called inertial range \citep{Frisch1995}. Previous observations show that the column density of atomic hydrogen (\ion{H}{1}) follows a power-law spatial power spectrum (SPS) extending from hundreds of parsecs down to a few tens of parsecs in the Milky Way \citep[e.g.][]{Green1993, Chepurnov2010, Mittal2023}, and from kiloparsec scales down to $\sim$100 pc in external galaxies \citep[e.g.][]{Nandakumar2020, Nandakumar2023}, while molecular gas traced by CO emission exhibits a similar scaling behavior over scales of a few to $\sim$10 pc down to sub-parsec scales (e.g. \citealt{Padoan2006, Sun2006, Pingel2018}). The velocity structure function (VSF) is a complementary measure, providing theoretical expectations from turbulence models such as Kolmogorov (K41 \citealp{Kolmogorov1941a}), She-Leveque (SL94 \citealp{She1994}), and Boldyrev (B02 \citealp{Boldyrev2002a}). Numerical and observational studies have explored these predictions \citep[e.g.][]{Kritsuk2007, Chira2019, Hu2022, Heyer2004}, and our previous work \citep{Ma2025} found scale-free VSFs within Galactic molecular clouds over $\sim$0.1-10 pc, with signatures of intermittency. Unlike \ion{H}{1} which pervades the entire Galactic disk, CO emission appears more like ``islands in an ocean,'' tracing discrete molecular clouds separated by diffuse inter-cloud regions. The VSF scalings traced by CO are often constrained to the cloud scale, leaving the turbulent scaling on intermediate scales between clouds and galactic structures poorly understood.

On galactic scales, spiral arms and bars can act as dynamical environments that regulate gas flows, where gas streaming motions and shocks can inject energy and interact with local turbulence (\citealt{Roberts1969, Kim2006, Meidt2013}). In this context, one may ask whether an entire spiral arm can be approximated as a coherent flow, and how such ordered motions connect to the random turbulent velocities measured within and between molecular clouds. Addressing how turbulence cascades from galactic structures down to star forming scales requires observations that bridge these intermediate scales. The Milky Way Imaging Scroll Painting (MWISP) survey \citep{Su2019, Yang2026}, one of the most comprehensive CO mapping programs of the northern Galactic plane in terms of sensitivity and coverage, provides such an opportunity.

In this work, we investigate how cloud-scale turbulence may arise from large-scale galactic flows. Using the high-sensitivity $^{12}$CO ($J$=1$-$0) data, we derive VSFs and SPSs from the large-scale maps and decompose the emission into individual molecular clouds to compute inter-cloud VSFs. Section \ref{sec2} describes the data and methods. Section \ref{sec3} presents the results. We discuss the results in Section \ref{sec4}, and summarize our conclusions in Section \ref{sec5}.

\section{Data and Methods} \label{sec2}
\subsection{Data and Slicing}\label{sec2.1}

The $^{12}$CO and $^{13}$CO data used in this work are part of the MWISP survey, 
which covers the Galactic longitude range $l=12^{\circ}$-$230^{\circ}$ and latitude $|b| \leq 5.25^{\circ}$ \citep{Su2019, Yang2026}. In this work, we focus on the Local Arm emission in the longitude range $l=105^{\circ}$-$150^{\circ}$. The $J=1$-0 transitions of $^{12}$CO, $^{13}$CO, and C$^{18}$O were observed simultaneously with the nine-beam SSAR receiver on the PMO 13.7 m telescope (\citealt{Shan2012}). The data have an angular resolution of $\sim$52$\arcsec$ ($\sim$0.25 pc at 1 kpc) and are sampled to pixels with sizes of 30$\arcsec\times30\arcsec$. The velocity resolution is 0.16-0.17 km s$^{-1}$, and the typical rms noise per channel is 0.5 K for $^{12}$CO and 0.3 K for $^{13}$CO and C$^{18}$O (\citealt{Cai2021}). Further details on the survey design, data quality, observing strategy, and full data release are provided in \citet{Su2019}, \citet{Cai2021}, and the recent MWISP data-release paper \citep{Yang2026}. 

\begin{figure*}[ht!]
\gridline{\fig{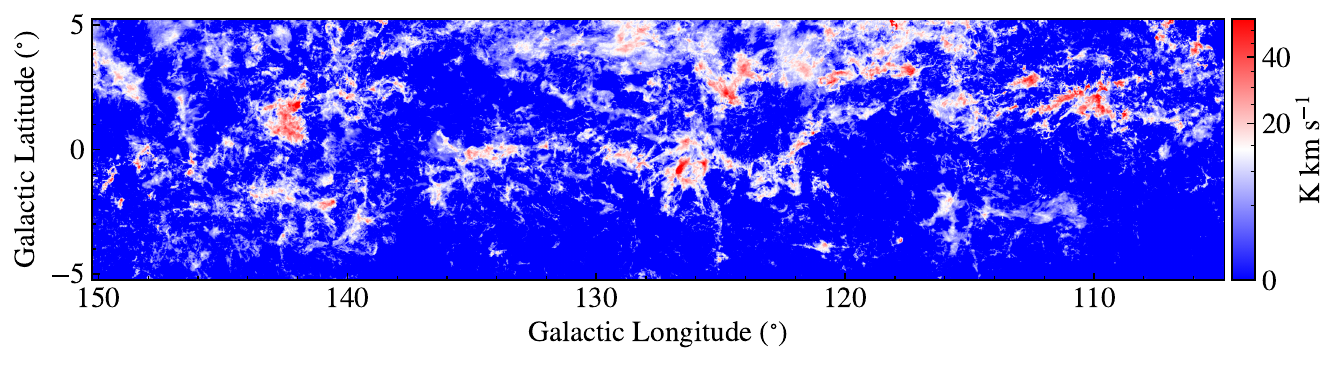}{0.97\textwidth}{(a)}}
\gridline{\fig{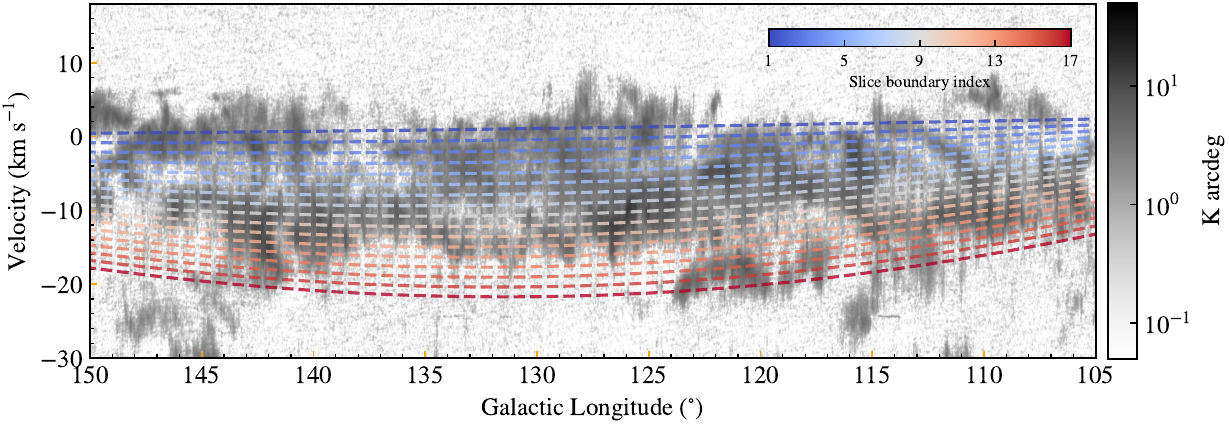}{0.99\textwidth}{(b)}}
\caption{(a) Integrated-intensity map of $^{12}$CO ($J$=1--0) emission in the selected Local Arm region at $l=105^{\circ}$--$150^{\circ}$, integrated over the velocity range from $-24$ to $10~\mathrm{km\,s^{-1}}$. (b) Longitude--velocity ($L$--$V$) diagram of the $^{12}$CO emission, integrated over the latitude range from $-5^{\circ}$ to $5^{\circ}$. The dashed curves indicate the longitude-dependent velocity boundaries of the slices derived from the Galactic rotation curve of \citet{Reid2019}. The colors denote the ordered slice boundaries.}
\label{fig1}
\end{figure*}

We aim to investigate the large-scale SFs and SPSs of the molecular gas in the Local Arm, as shown in the integrated-intensity map within $-24$ to $10~\mathrm{km\,s^{-1}}$ (Figure~\ref{fig1}(a)). To ensure that the analyzed emission traces physically related gas and to minimize the influence of the large-scale velocity pattern associated with Galactic rotation, we decompose the PPV cube into 17 narrow velocity slices that follow the longitude-dependent $L$--$V$ structure.

The slice boundaries are defined by loci derived from the Galactic rotation curve of \citet{Reid2019}, calculated using the ``Kinematic Distance Utilities'' package \citep{Wenger2018}, and are shown in Figure~\ref{fig1}(b). The slices are indexed in order of increasing nominal kinematic distance rather than labeled by distance values. This is because the Local Arm lies close to the solar circle, where kinematic distances are intrinsically uncertain, and it is an extended structure with a finite line-of-sight depth \citep{Xu2013, Reid2019}. Each velocity slice spans approximately $1$--$2~\mathrm{km\,s^{-1}}$. The molecular emission changes smoothly from one slice to the next, indicating that the slices sample adjacent parts of contiguous structures in PPV space (see Appendix~\ref{app:slice_maps}). Besides, extinction--parallax measurements indicate that many of the molecular clouds associated with the large-scale coherent emission in this longitude range lie at distances of approximately 0.8$-$1.2~kpc \citep{Yan2021b}. We therefore adopt a common fiducial distance of 1 kpc for all slices when converting angular scales into physical scales. This value is intended to provide a representative physical scale for the Local Arm segment analyzed here, rather than a precise distance to every emitting structure.

For each velocity slice, we derive the intensity weighted velocity map and the integrated intensity map. Before the calculation of VSFs and V-SPSs, the large-scale velocity trend associated with Galactic rotation is removed. A smooth velocity surface is constructed to represent the systematic gradient implied by the kinematic-distance structure (dashed lines in Figure \ref{fig1}(b)). This surface is subtracted from the centroid-velocity map so that the resulting statistics reflect only turbulent fluctuations rather than global shear. 

\subsection{Statistical Methods} \label{sec2.2}
\subsubsection{Velocity Structure Function} \label{sec2.2.1} 
The p-th order VSF is defined as
\begin{equation}
    \label{eq1}
    S_p(\mathbf{l}) = \langle |\mathbf{v}_{\mathbf{x}} - \mathbf{v}_{\mathbf{x+l}}|^p \rangle,
\end{equation}
where $\mathbf{v}$, $\mathbf{x}$, and $\mathbf{l}$ are the velocity, position, and spatial lag vectors, respectively. The theoretical expectations for the VSF scaling exponents are summarized in \citet{Ma2025}; here we describe only the computation procedure. The velocity map contains $\sim$2.7$\times10^6$ pixels, and to reduce memory usage we adopt an image-shift method rather than explicitly selecting non-repeating pixel pairs. For a given direction, the velocity map is shifted from 2 pixels to half the long edge of the region (2700 pixels). The final VSF is obtained by averaging over all sampled directions, and the uncertainty is taken as the standard deviation among them. In this work, we uniformly sample the angular directions from $-15^\circ$ to $15^\circ$ with a step of $1^\circ$. The angular range is restricted because the mapped region is highly elongated in Galactic longitude, and sampling directions close to the $L$-axis maximizes the available lag range while avoiding edge truncation.

\subsubsection{Spatial Power Spectrum} \label{sec2.2.2}

The power spectrum (PS) is defined as the Fourier transform of the autocorrelation function (ACF) of an observable A,
\begin{equation}
    \label{eq2}
    P(\mathbf{k}) =  \iint{ \langle A(\mathbf{x}) A(\mathbf{x+l})\rangle e^{-i\mathbf{k}\cdot\mathbf{l}} d\mathbf{l},}
\end{equation}
where A can be either the velocity or the intensity of the molecular gas, $\mathbf{k}$ is the wave number vector, and $\mathbf{l}$ is the spatial lag vector. In practice, we compute the spatial power spectrum (SPS) by taking the 2D Fourier transform of each velocity or intensity map and evaluating the squared amplitudes in elliptical annuli in the Fourier space, accounting for the rectangular map geometry. To reduce edge effects, we apply an apodization procedure to the molecular-cloud boundary \citep[e.g.,][]{Pingel2018}. Specifically, the mask of the emitting region (1 inside, 0 outside) is convolved with a Gaussian kernel of $\sigma = 2$ pixels to produce a tapered boundary that gradually falls to zero, and the resulting mask is multiplied with each map prior to the FFT. After FFT, the azimuthal median amplitude within each annulus is adopted as the PS value at that wave number. The wave number $k$ is defined as $k = \sqrt{u^2 + v^2} = 2\pi / \lambda$, where $u$ and $v$ are the Fourier space coordinates, and $\lambda$ is the spatial wavelength of interest. 

We further test the statistical isotropy of the VSF and SPS measurements and assess the influence of observational noise using emission-free velocity channels. These tests are presented in Appendix~\ref{app:validation_tests}. The results demonstrate that the VSF and SPS measurements are statistically isotropic and that the noise does not significantly affect the measured scaling.

\section{Results} \label{sec3}
\subsection{Structure Function Scaling} \label{sec3.1}
\begin{figure*}[htb!]
\gridline{\fig{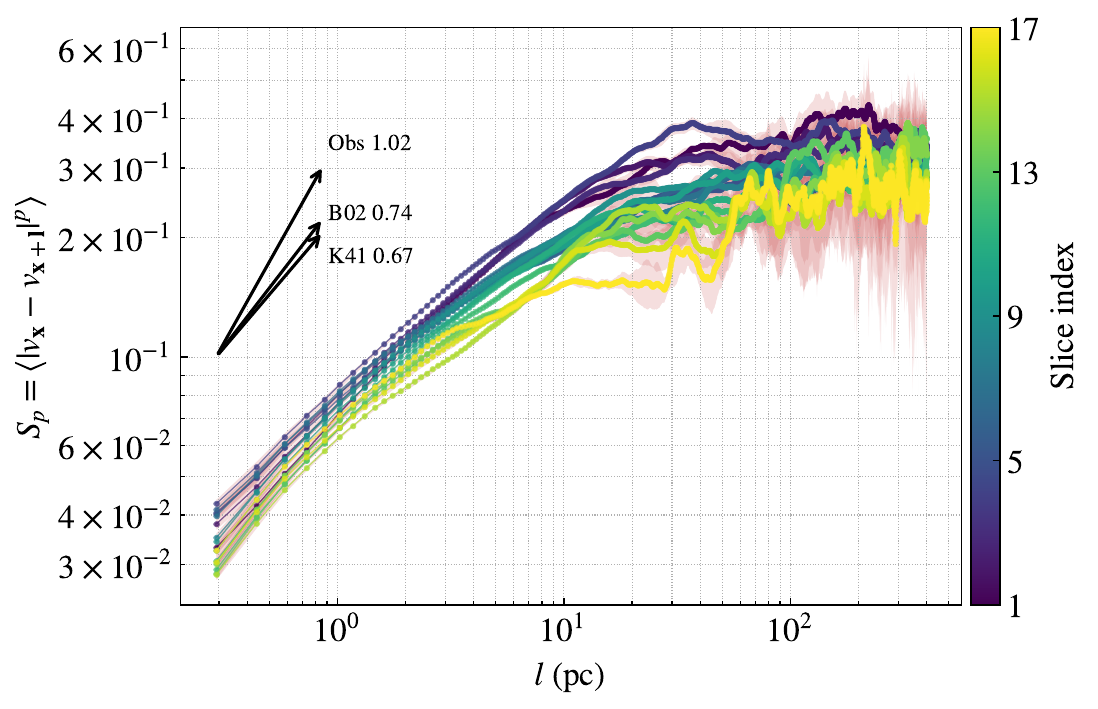}{0.5\textwidth}{(a)}
          \fig{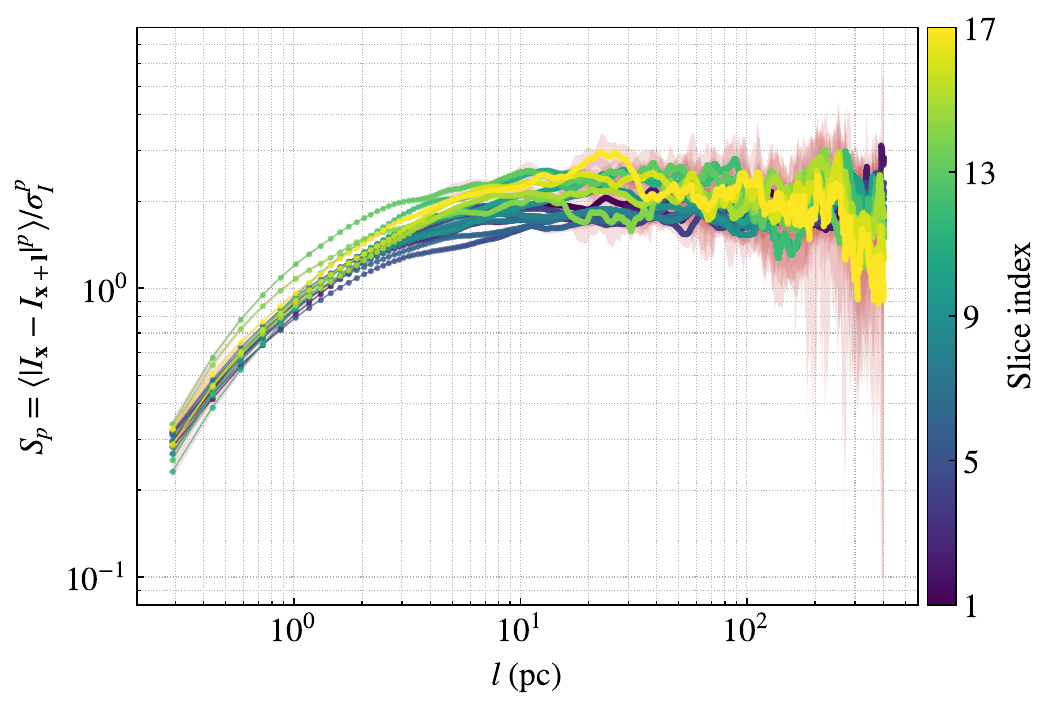}{0.48\textwidth}{(b)}}
\caption{Second-order (p=2) SFs of (a) the gradient-corrected intensity-weighted velocity (moment 1) and (b) the integrated intensity (moment 0) for different slices. The color of each curve indicates the slice index. The theoretical slopes from the \citetalias{Kolmogorov1941a} and \citetalias{Boldyrev2002a} models, as well as the average slope for individual molecular clouds from \citet{Ma2025}, are shown as arrows for reference.}
\label{fig2}
\end{figure*}

Figure \ref{fig2}(a) presents the second-order VSFs calculated for different velocity slices, color-coded by their slice indices. The VSFs consistently follow a power-law scaling over nearly two decades in scale, with an observed slope of $\sim$0.65. This slope is close to the theoretical prediction from the \citetalias{Kolmogorov1941a} model, being only slightly shallower. A notable feature is that the scaling saturates around 15-30 pc (50--100$^{\prime}$), where the VSFs become flat above this scale interval. The overall shapes of the VSFs are quite similar across different velocity layers. A power-law VSF over a finite range of scales is widely regarded as a key statistical signature of turbulence in the ISM. Therefore, the power-law behavior observed in Figure \ref{fig2}(a) indicates that the molecular gas traced by $^{12}$CO in this portion of the Local Arm is turbulent at least below $\sim$30 pc. As demonstrated by the sampling tests in Section \ref{sec4.2}, the flattening of the VSFs is more naturally explained by the sparse sampling of the CO emission than by a universal energy injection scale. In Figure \ref{fig1}(a), the angular separation of $\sim$50--100$^{\prime}$ (0.8--1.7$^{\circ}$) roughly matches the typical spacing between major emission peaks in the Local Arm. Therefore, the flattening of the VSFs above 15-30 pc likely reflects the limited spatial extent of coherent molecular structures. In other words, the turbulent cascade traced by the $^{12}$CO VSFs appears to be confined to scales comparable to individual molecular clouds.

For the intensity SFs, their absolute values differ among layers because of the varying column densities. Therefore, we normalize each SF by dividing it by the standard deviation of the corresponding intensity map. After this normalization, as shown in Figure \ref{fig2}(b), the intensity SFs exhibit similar shapes across different velocity slices, indicating that the column-density fluctuations in different layers are governed by the same underlying physical processes. They do not show any clear power-law scaling but instead display a smooth curvature at scales below $\sim$50--100$^{\prime}$ and saturate above that scale. In compressible molecular clouds, intensity statistics can be strongly influenced by projection effects, radiative transfer, and physical processes such as intermittency, shocks, and gravity, which can obscure the underlying scale-free behavior. Therefore, the absence of a clear power-law scaling in the intensity SFs is not unexpected and does not necessarily contradict the presence of a turbulent cascade inferred from the velocity statistics. The saturation scale likely reflects the characteristic size of coherent molecular structures within each velocity slice.

\subsection{Spatial Power Spectrum Scaling} \label{sec3.2}
\begin{figure*}[htb!]
\gridline{\fig{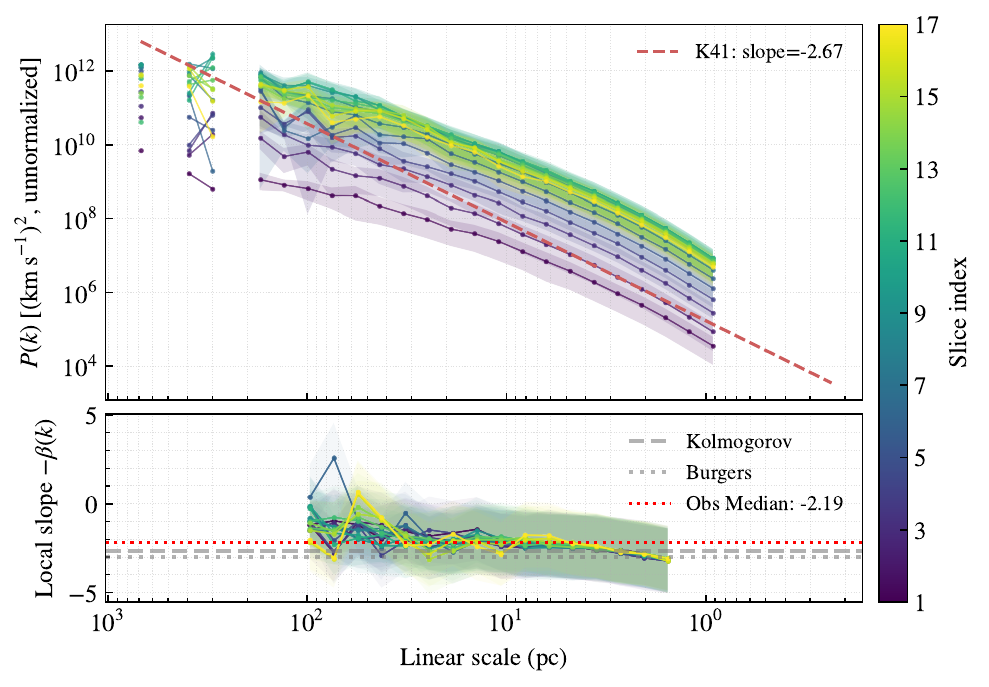}{0.5\textwidth}{(a)}
          \fig{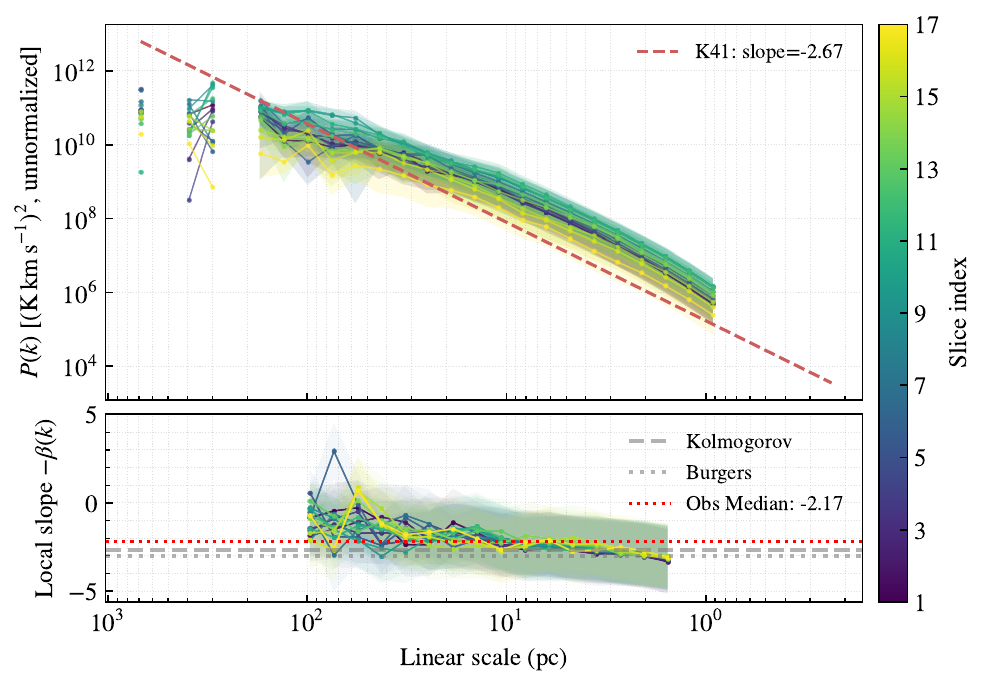}{0.5\textwidth}{(b)}}
\caption{SPSs of (a) the intensity-weighted velocity and (b) the integrated intensity for different velocity slices. The color of each curve indicates the slice index, as in Figure~\ref{fig1}b. The theoretical slopes from the \citetalias{Kolmogorov1941a} and Burgers \citep{bec2007burgers} models are shown as dashed lines.}
\label{fig3}
\end{figure*}

Figure \ref{fig3} presents the velocity and intensity SPSs for different velocity slices. In contrast to the VSFs, the SPSs exhibit an approximately scale-free form over a broad range of spatial scales, from $\sim$1 to $\sim$200 pc. The sparsely sampled larger-scale points ($\sim$ 400 pc) are broadly consistent with an extrapolation of the same trend. At scales smaller than about 10 pixels ($\sim$1 pc at 1 kpc), the SPSs are affected by the apodization process before the calculation, while at large scales ($>\sim$200 pc), they are influenced by the limited number of pixels in the Fourier space. We restrict the quantitative analysis to spatial scales between 1 and 200 pc, excluding the smallest scales affected by apodization and the largest scales with limited Fourier sampling. Rather than assigning a single power-law exponent to each SPS, we use a moving-window fit with a width of four bins to characterize its scale-dependent local slope, as shown in Figure \ref{fig3}. The lower panels display the resulting local slopes. Although the slopes show a mild tendency to become more negative toward smaller spatial scales, the variation is gradual and no distinct spectral break is present over the fitted range. The local slopes of the velocity SPSs span approximately $-1$ to $-3$, with a median value of $\sim-2.19$, which is slightly shallower than the \citetalias{Kolmogorov1941a} prediction of $-8/3$ ($-2.67$). The intensity SPSs have a median local slope of $\sim-2.17$ and span a similar range ([$-$3, $-$1]). Both the velocity and intensity SPSs therefore retain smooth, approximately scale-free behavior across the sampled scales, consistent with a continuous turbulent cascade rather than a transition between two distinct scaling regimes. 

From the above results, although the VSFs exhibit power-law scaling only over a limited range at small scales, the flattening of the SFs does not correspond to the actual energy injection scale of the turbulence, since both the velocity and intensity SPSs reveal a continuous cascade extending over nearly the entire range of spatial scales. This does not imply that the SF analysis is meaningless; rather, the SFs capture the scaling behavior at the smallest scales, where the SPSs are affected by the Gaussian tapering applied to the boundaries of the molecular emission.

To interpret the connection between these diagnostics, it is useful to recall their mathematical relationships. The PS and the ACF form a Fourier-transform pair, while the SF can be expressed in terms of the ACF. Under ideal conditions, the power-law indices of the SPSs and the second-order SFs are related by $m = n - 2$, where $n$ and $m$ are the SPS and SF exponents ($P(k)\propto{k^{-n}}$), respectively. However, this correspondence holds only within limited ranges of spectral indices and scaling intervals \citep{Hou1998}. As shown by \citet{Hou1998}, the convergence of the SF and the ACF depends on the slope of the underlying power spectrum, such that SFs can reliably recover the power-spectrum slope only within a restricted range of spectral indices (corresponding to SPS slopes between 2 and 4). In our analysis, the SPS slopes vary from $\sim -1$ at large scales to $\sim -3$ at small scales. At large scales, the inferred spectral indices therefore fall outside the range where SFs can reliably recover the power-spectrum scaling. As a result, the apparent scaling range inferred from VSFs is expected to be shorter than that inferred from SPSs. This behavior is a general property of turbulence statistics and has also been discussed in \citet{Frisch1995}. It naturally explains why the VSFs in our analysis exhibit power-law behavior only below a certain scale, while the SPSs display power-law scaling over a much broader range of spatial scales.

The SPS slopes measured in this work are broadly consistent with previous measurements in both external galaxies and the Milky Way. CO and H{\sc i} observations of external galaxies often report two-dimensional SPS slopes of approximately $-$1.5 to $-$2 \citep[e.g.,][]{Dutta2009, Nandakumar2020}, while Galactic ISM studies find slopes ranging from about $-$2.6 to $-$4 across different tracers and environments \citep[e.g.,][]{Pingel2018}. The slopes obtained here therefore fall within the range commonly observed for turbulent ISM structure, supporting the interpretation that the $^{12}$CO emission in this portion of the Local Arm traces a turbulent cascade over a wide range of spatial scales.

\subsection{Extended Self-similarity Scaling across Individual Molecular Clouds} \label{sec3.3}

\begin{figure*}[htb!]
\centering
\includegraphics[width=0.6\textwidth]{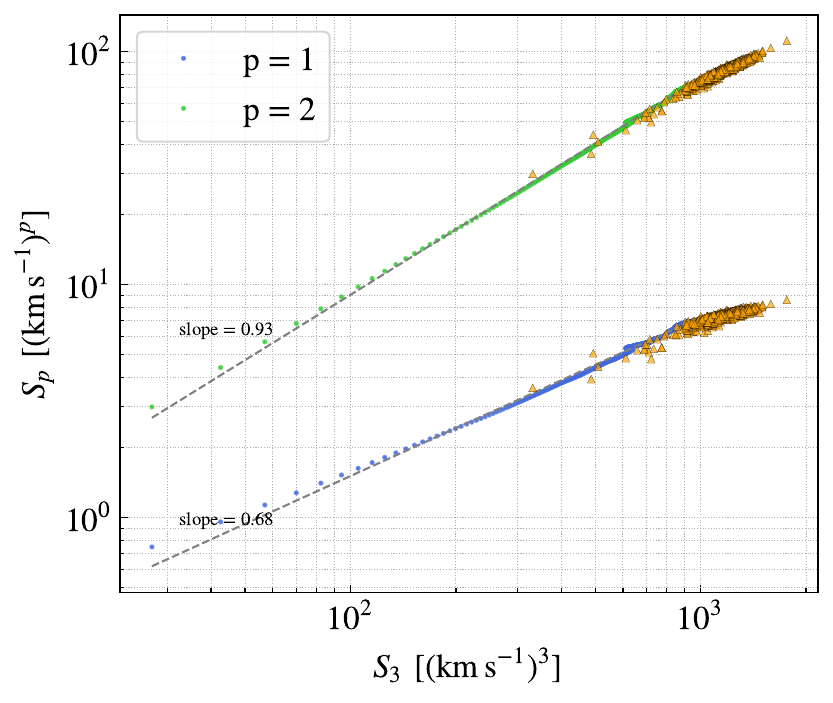}
\caption{Extended self-similarity (ESS) scaling analysis, showing the relationship between S$_i$ (for $i = 1, 2$) and S$_3$ for the molecular gas. The dots represent the VSFs calculated pixel by pixel from the $^{12}$CO centroid velocity map, while the orange triangles denote the VSFs derived from the published MWISP $^{12}$CO molecular-cloud catalog of \citet{Yan2021a}. The dashed lines indicate the best-fit power-law relations for the VSFs of the velocity map derived within $[-24, 10]$ km s$^{-1}$. The corresponding power-law exponents are labeled above each dashed line.}
\label{fig4}
\end{figure*}
From the previous section, we have established that the turbulent cascade traced by $^{12}$CO emission extends over a wide range of spatial scales, from sub-parsec (from VSFs) to hundreds of parsecs (from SPSs). Different velocity slices within this portion of the Local arm exhibit similar VSFs, intensity SFs, velocity SPSs, and intensity SPSs. We could make a reasonable assumption that the Local arm can be treated as a continuous medium with similar turbulence cascading properties across its extent. Hence, we can use the entire velocity range of $-$24 to 10 km s$^{-1}$ to compute the VSFs and examine the nature of the velocity difference between individual molecular clouds. 

Here we compare VSFs derived from the centroid-velocity map ($-$24 to 10 km s$^{-1}$) with cloud-to-cloud VSFs derived from the published MWISP $^{12}$CO molecular-cloud catalog of \citet{Yan2021a}. The catalog was constructed with DBSCAN (\citealt{Ester1996}), which identifies spatially and kinematically connected CO-emitting structures in PPV space. We select cataloged clouds within the same Local Arm region and use their centroid positions and velocities in the VSF calculation. However, unlike the velocity slices in Section \ref{sec3.2}, the full Local Arm emission occupies a relative broad band in the $l$-$v$ diagram, so a unique Galactic-rotation correction for the whole arm is not well defined and would be model-dependent, especially for the discrete cloud catalog. Therefore, we do not subtract a global Galactic-rotation model from the centroid-velocity map or from the cloud catalog.

The extended self-similarity (ESS; \citealt{Benzi1993}) is a technique that enhances the effective scaling range of SFs by plotting them against the third-order SF, rather than against spatial separation. This approach has been widely used in turbulence studies to extract scaling exponents when a clear inertial range cannot be identified from conventional SFs alone. The physical basis of ESS has been investigated in subsequent theoretical work. \citet{Fujisaka2001} showed that ESS naturally arises when SFs of different orders share a common crossover-modified scaling variable, allowing robust scaling relations to persist even outside a well-developed inertial range. More recently, \citet{Chakraborty2010} conducted a theoretical analysis of SFs in a solvable turbulence model, providing a rigorous demonstration of ESS.

In our case, the second-order VSFs derived from the velocity maps exhibit power-law scaling only over a limited range of small scales. Therefore, ESS provides a natural framework to test whether velocity fluctuations on different physical scales share the same underlying scaling relations. To examine whether the velocity differences between individual molecular clouds follow the same scaling behavior as those within clouds, we compute VSFs for the identified $^{12}$CO molecular clouds, treating each cloud as a PPV particle characterized by its centroid position and velocity. Before the calculation, we exclude clouds that overlap with others within their projected boundaries, so that only velocity differences between distinct clouds are considered. By contrast, the VSFs derived from the centroid velocity map include both intra-cloud velocity fluctuations and inter-cloud bulk motions at comparable spatial scales. The comparison between these two measurements within the ESS framework, therefore, allows us to directly test whether cloud-to-cloud motions are consistent with an extrapolation of the same turbulent cascade inferred from the internal velocity field.

Figure \ref{fig4} displays the ESS scaling relations for both the continuous velocity map (dots) and the discrete molecular clouds (triangles). The VSFs from the continuous map exhibit clear power-law relations when plotted against S$_3$, with best-fit exponents of 0.68 for S$_1$ vs. S$_3$ and 0.93 for S$_2$ vs. S$_3$. Remarkably, the VSFs derived from the discrete molecular clouds align well with those from the continuous map, following the same power-law trends. This indicates that the velocity differences between individual molecular clouds adhere to the same scaling relations as those observed within clouds. 

\section{Discussion} \label{sec4}
\subsection{Influence of Sampling on the VSF and SPS Statistics} \label{sec4.2}

\begin{figure*}[htb!]
\gridline{\fig{{im5.0}.pdf}{0.45\textwidth}{(a)}
\fig{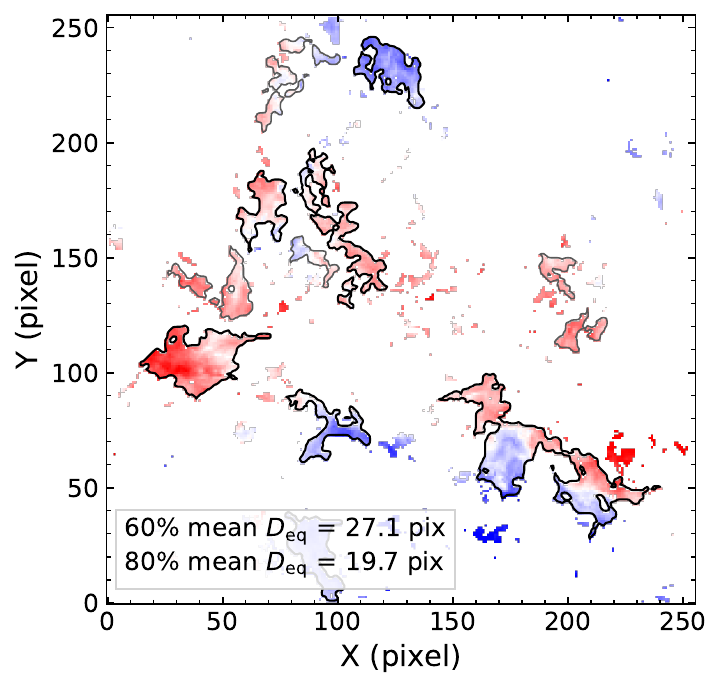}{0.45\textwidth}{(b)}}
\gridline{\fig{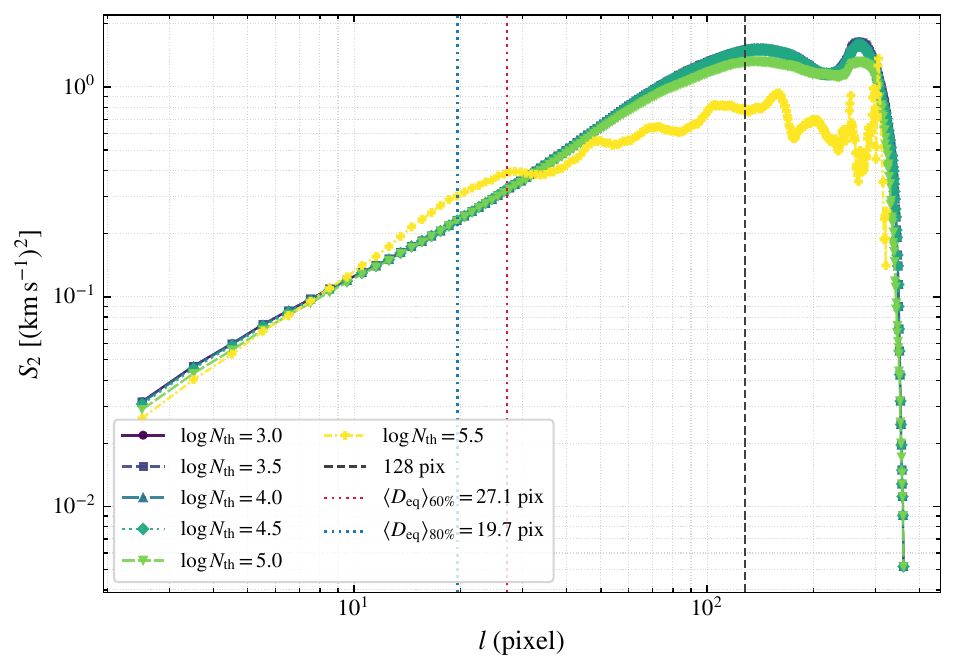}{0.45\textwidth}{(c)}
 \fig{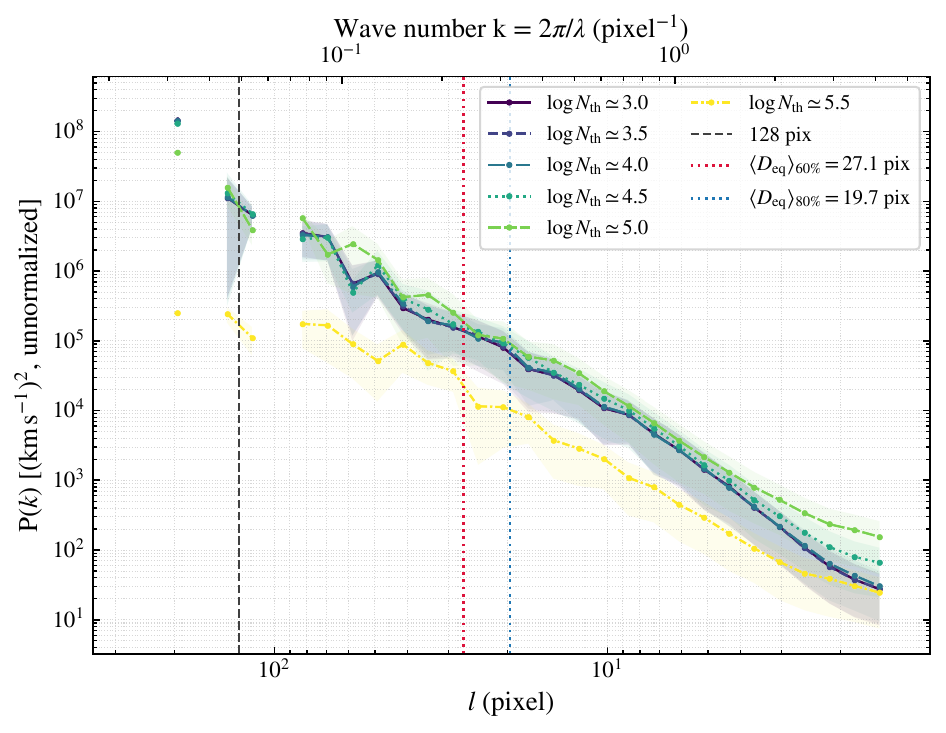}{0.44\textwidth}{(d)}}
\caption{Tests of sampling effects on the VSFs and SPSs using MHD simulation data. (a) and (b): Examples of masks generated using threshold values of $10^{5}$ and $10^{5.5}$, respectively. In panel (b), black contours enclose the connected patches that cumulatively account for the first 60\% of the total masked area, while gray contours indicate the additional patches required to reach 80\% of the total masked area. The mean equivalent diameters of the patches in the 60\% and 80\% cumulative-area samples are 27.1 and 19.7 pixels, respectively. (c) and (d): The resulting second-order VSFs and SPSs computed from the masked velocity fields. Different colors correspond to different masking thresholds. The vertical dotted lines mark the two characteristic patch sizes, while the vertical dashed line marks the half the full box scale.}
\label{fig5}
\end{figure*}

In Section \ref{sec3.1}, we showed that the VSFs exhibit a more limited power-law scaling range than the SPSs, while the physical origin of the turnover remains unclear. To investigate this effect, we carried out a set of tests using MHD simulation data to assess how spatial sampling influences VSF and SPS statistics. We adopted the supersonic MHD simulation of \citet{Mocz2017} at $t=0$, when turbulence is fully developed and self-gravity is still turned off. The model has a uniform magnetic field of $B=10~\mu$G, a sonic Mach number of $\mathcal{M}_{\rm s}=10$, and a resolution of $256^3$.

From the simulation cube, we constructed a mass-weighted velocity map and a ``column-density'' map by averaging and summing along the $Z$ axis. We then applied a series of column-density thresholds to generate masks that progressively reduce spatial sampling, and used these masks to restrict the velocity field. Figures \ref{fig5}(a) and \ref{fig5}(b) show two representative examples.

As shown in Figures \ref{fig5}(c) and \ref{fig5}(d), decreasing spatial sampling has different effects on the VSFs and SPSs. The VSFs for the lower thresholds nearly overlap because these masks retain most of the spatial area. A clear break appears only for the highest threshold, where the mask becomes fragmented into discrete emitting patches. By contrast, the SPSs show only local fluctuations near the corresponding spatial scales and a reduction in overall power, while their large-scale power-law trends are largely preserved.

To quantify the characteristic sampling scale, we identify the connected emitting patches in the highest-threshold mask and sort them by area. We then compute the mean equivalent diameter, $D_{\rm eq}=2\sqrt{A/\pi}$, of the largest patches whose cumulative area reaches 60\% and 80\% of the total masked area. These characteristic sizes are $\langle D_{\rm eq}\rangle_{60\%}=27.1$ pixels and $\langle D_{\rm eq}\rangle_{80\%}=19.7$ pixels, respectively. As marked in Figure \ref{fig5}(c), these values corresponds to the scale at which the highest-threshold VSF departs from the small-scale power-law trend. This correspondence indicates that the VSF turnover is controlled by the characteristic size of the connected emitting regions. The comparison with the SPSs further shows that this turnover is not caused by a change in the underlying turbulent cascade of the original velocity field. Instead, limited spatial sampling can produce an apparent break in the VSFs while the SPS power-law slope remains largely unchanged. We therefore conclude that the large-scale flattening of the MWISP VSFs is likely a consequence of the discrete and sparse distribution of molecular clouds traced by $^{12}$CO emission.

\subsection{Implications for Molecular Cloud Formation and Star Formation}\label{sec4.4}

Numerical simulations and observational studies have shown that molecular clouds are assembled from diffuse atomic and molecular gas through converging flows driven by galactic-scale processes such as spiral arm compression, shear, and large-scale streaming motions \citep[e.g.][]{Dobbs2014, Fujii2021}. These processes set the initial physical and kinematic conditions of molecular clouds during their assembly, well before star formation becomes dominant. From a dynamical point of view, \cite{Meidt2018} proposed that molecular clouds should be understood as density enhancements embedded within a continuous galactic velocity field, rather than as independent turbulent entities. Complementary observational evidence for this view was presented by \cite{Henshaw2020}, who combined multi-scale datasets to identify characteristic length scales associated with cloud formation and demonstrated that molecular clouds are embedded within nested, interdependent gas flows across a wide range of spatial scales. 

In the present study, we extend these results by directly tracing turbulence statistics across a continuous range of spatial scales using a single, homogeneous dataset, without decomposing the emission into individual clouds. The consistency between cloud-based and large-scale velocity structure functions demonstrates that the apparent universality of cloud-scale turbulence arises from the inheritance of velocity correlations from larger scales. Recent reviews have emphasized that the connection between galactic-scale and cloud-scale star formation is mediated through the lifecycle of molecular clouds, with galactic dynamics regulating cloud formation rates and lifetimes, while local processes modulate star formation within individual clouds \citep[e.g.][]{Chevance2023}. Viewed in this context, our results suggest that star formation across scales should be understood as a continuous dynamical process, rather than as separate phenomena on galactic and cloud scales.

\section{Conclusion} \label{sec5}

Using $^{12}$CO (J=1--0) data from the MWISP survey, we have analyzed the turbulent properties of molecular gas in a $\sim$kpc-scale segment of the Local Arm by decomposing the PPV cube into narrow velocity slices that follow the Galactic-rotation trend in the $L$-$V$ diagram. 
We find that both the SFs and SPSs exhibit nearly identical scaling behaviors across all velocity slices, demonstrating that the molecular gas belongs to a single, coherent turbulent field extending continuously from $\sim 200$~pc down to 1~pc scales, while the sparsely sampled velocity SPS at larger scales (up to $\sim 400$~pc) remains broadly consistent with an extrapolation of the same trend.  

By further comparing pixel-based turbulence statistics with cloud-to-cloud VSFs derived from a discrete molecular-cloud catalog, we show that inter-cloud motions follow the same turbulent cascade traced by the continuous medium. This indicates that molecular clouds are not dynamically isolated entities, but instead represent localized density enhancements embedded within a larger-scale turbulent flow. Taken together, these results provide direct observational evidence that the dynamical state of molecular clouds is inherited from galactic-scale motions, offering a natural bridge between galactic dynamics and star formation on cloud scales.

\begin{acknowledgments}
    We thank the anonymous referee for the constructive comments that help to improve this manuscript. This research made use of the data from the Milky Way Imaging Scroll Painting (MWISP) project, which is a multi-line survey in $^{12}$CO/$^{13}$CO/C$^{18}$O along the northern galactic plane with PMO-13.7 m telescope. We are grateful to all the members of the MWISP working group, particularly the staff members at PMO-13.7 m telescope, for their long-term support. MWISP was sponsored by National Key R$\&$D Program of China with grants 2023YFA1608002 $\&$ 2017YFA0402701 and by CAS Key Research Program of Frontier Sciences with grant QYZDJ-SSW-SLH047. Y.M. acknowledges the support of NSFC grant 12303033. H.W., Y.M., and M.Z. acknowledge the support of NSFC grant 12473026. X.C. acknowledges the support of NSFC grant 12041305 and the support from the Tianchi Talent Program of Xinjiang Uygur Autonomous Region. M.Z. acknowledges the support of NSFC grant 12073079.
\end{acknowledgments}

\section{Appendix information}

\appendix

\section{Representative Maps of the Velocity Slices}\label{app:slice_maps}

\begin{figure*}[htb!]
\centering
\includegraphics[width=0.95\textwidth]{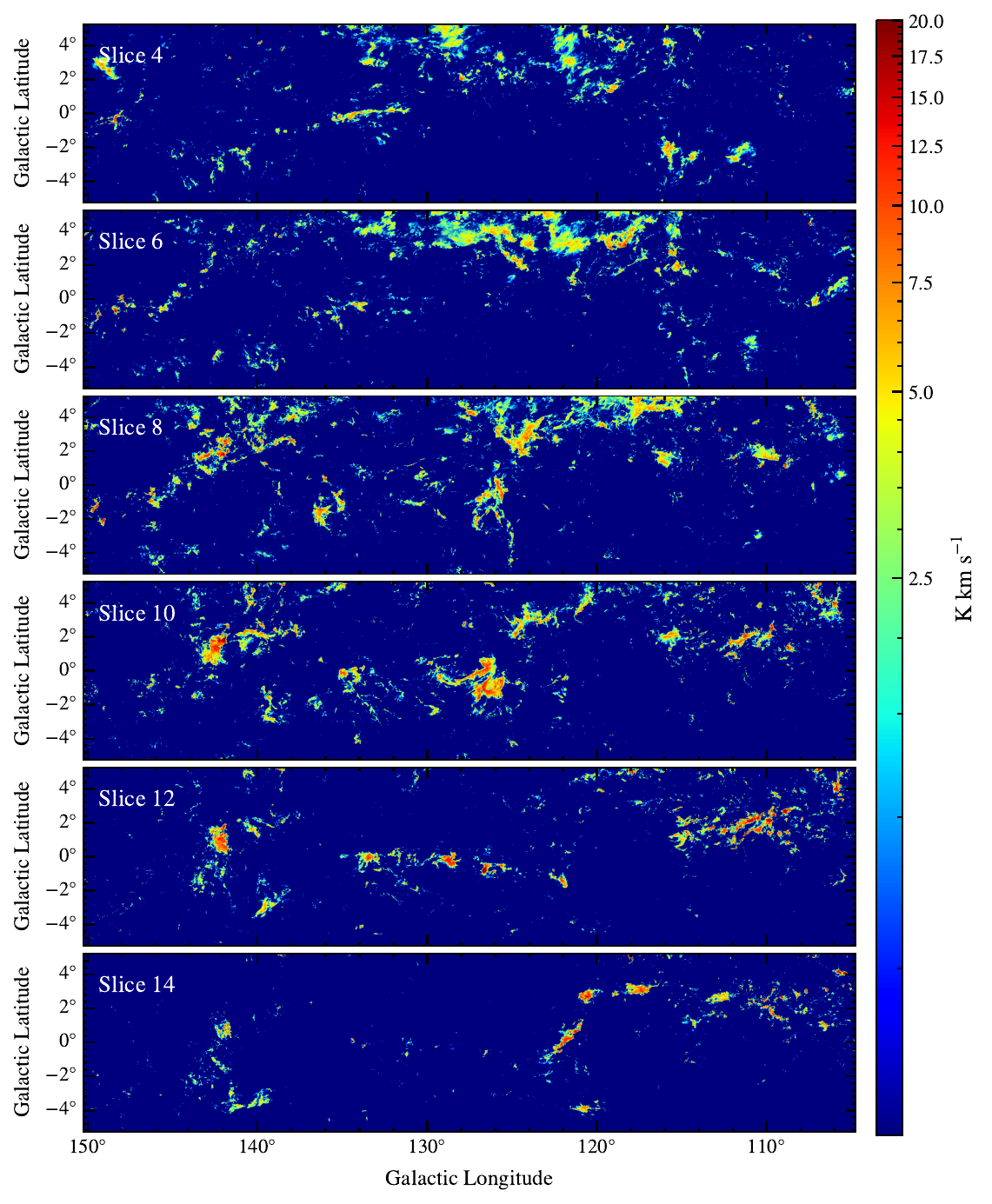}
\caption{Representative integrated-intensity maps from six velocity slices used in the analysis. The panels are ordered following the sequence of slices shown in Figure \ref{fig1}(b).}
\label{fig:app_slice_maps}
\end{figure*}

\section{Validation Tests for Noise Effects and Statistical Isotropy}
\label{app:validation_tests}

\subsection{Statistical-isotropy Test}
\label{app:isotropy_test}

The full MWISP map analyzed in this work has a large aspect ratio, extending over $45^{\circ}$ in Galactic longitude but only $10.5^{\circ}$ in Galactic latitude. In the main analysis, the VSFs are therefore calculated using directions close to the Galactic-longitude axis in order to maximize the accessible lag range and avoid severe edge truncation. To test whether this directional restriction biases the inferred scaling, we selected two square subregions, each covering $8^{\circ}\times8^{\circ}$, and repeated the second-order VSF calculation for position angles from $0^{\circ}$ to $179^{\circ}$ with a step of $1^{\circ}$. The square geometry allows all directions to be sampled over the same range of angular separations and therefore provides a direct test of statistical isotropy on scales smaller than the field size.

\begin{figure*}[htb!]
\gridline{\fig{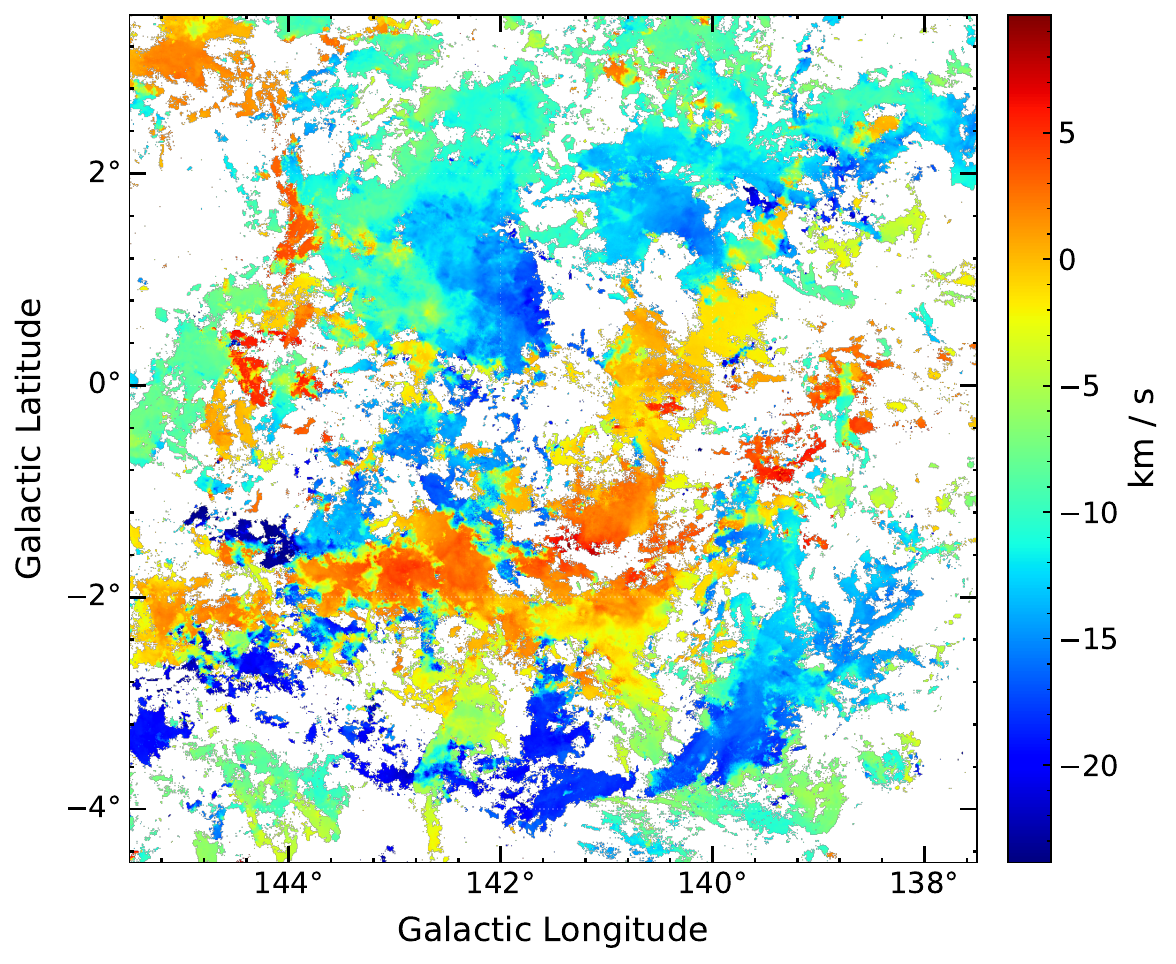}{0.48\textwidth}{(a)}
          \fig{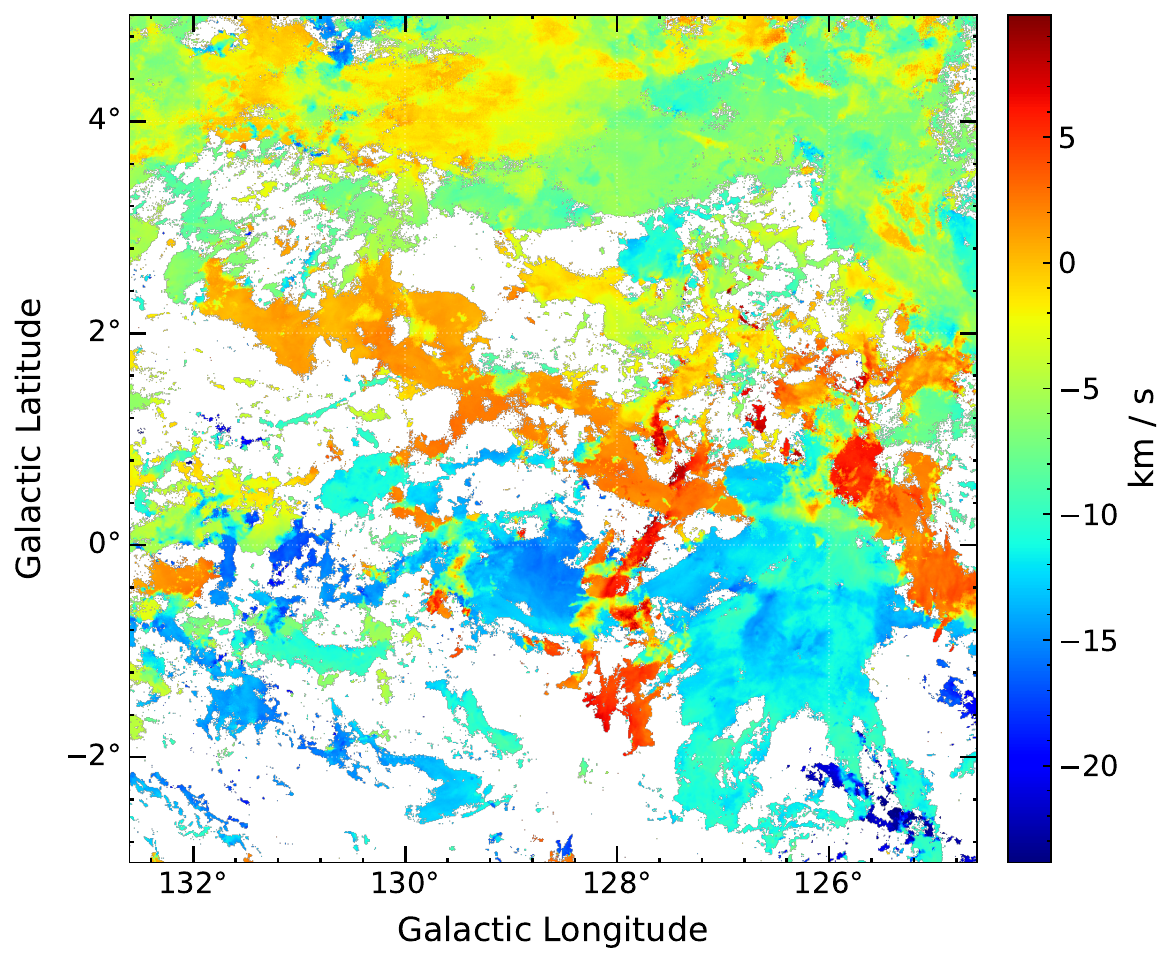}{0.48\textwidth}{(b)}}
\caption{The two $8^{\circ}\times8^{\circ}$ centroid-velocity subregions used for the statistical-isotropy test. Their approximately square geometry allows a common range of angular separations to be sampled over all position angles.}
\label{fig:app_isotropy_regions}
\end{figure*}

\begin{figure*}[htb!]
\gridline{\fig{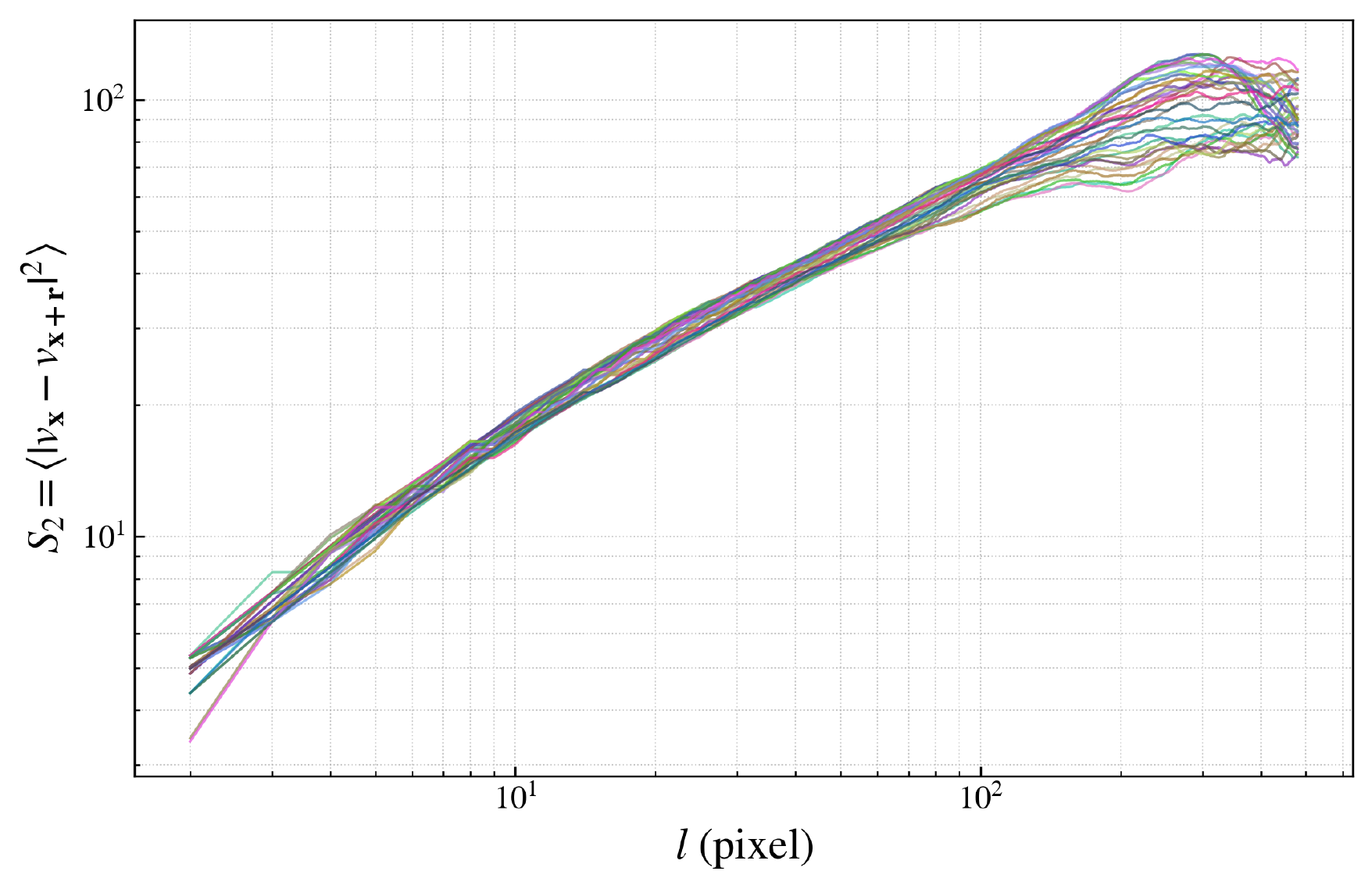}{0.48\textwidth}{(a)}
\fig{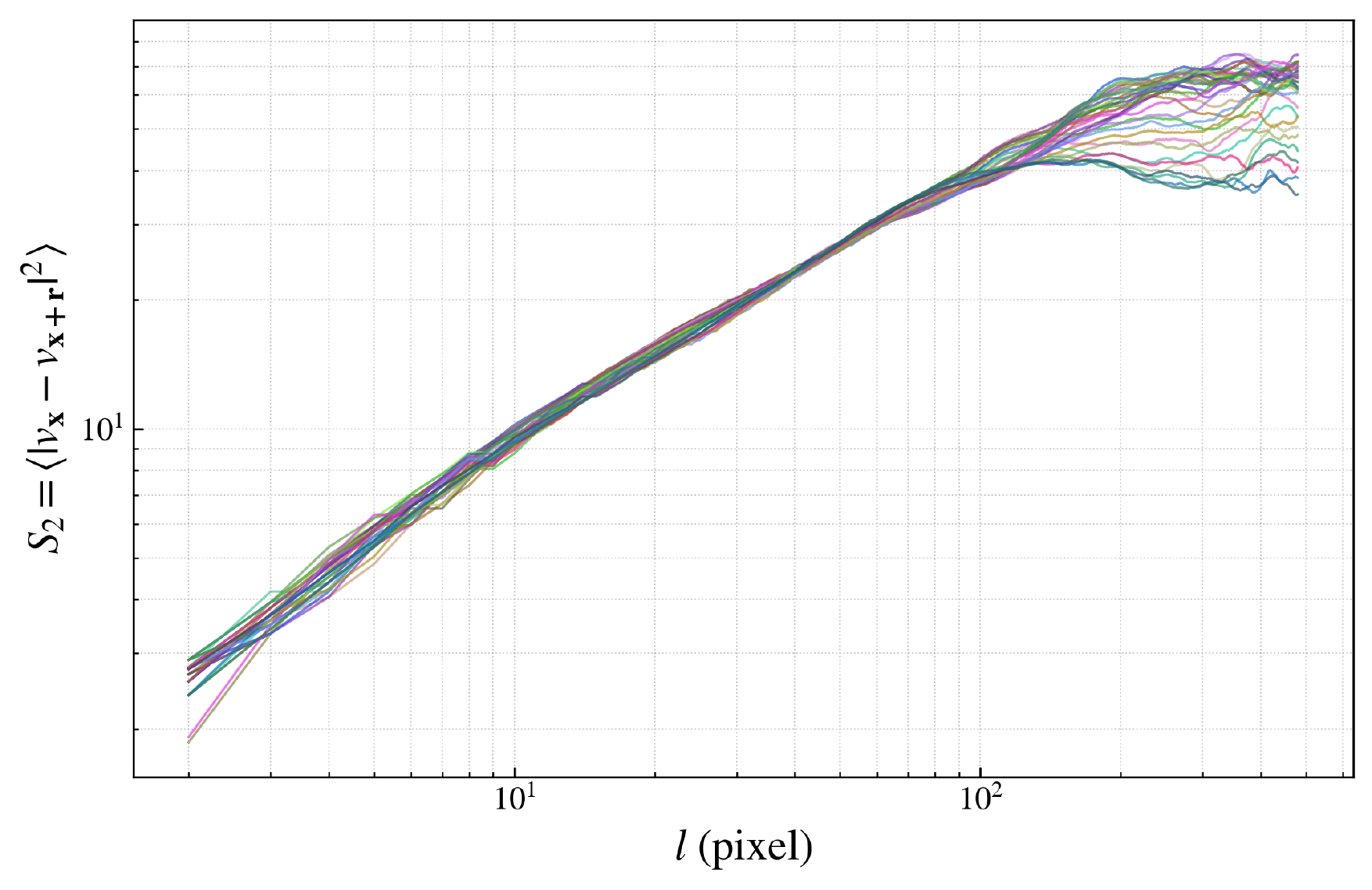}{0.48\textwidth}{(b)}}
\caption{Directional second-order VSFs for the two subregions shown in Figure~\ref{fig:app_isotropy_regions}. The VSFs are calculated for position angles from $0^{\circ}$ to $179^{\circ}$ in steps of $1^{\circ}$.}
\label{fig:app_isotropy_vsf}
\end{figure*}

Figure~\ref{fig:app_isotropy_regions} shows the two subregions used in this test, and Figure~\ref{fig:app_isotropy_vsf} presents the corresponding directional VSFs. For both subregions, the VSFs measured at different position angles closely overlap throughout the common lag range. We detect no systematic azimuthal dependence in either the VSF amplitude or its scaling behavior. The velocity statistics are therefore consistent with isotropy on the scales that can be fully sampled within an $8^{\circ}\times8^{\circ}$ field. Hence, restricting the main analysis to directions near the Galactic-longitude axis has only a limited effect on the inferred VSF scaling.

\begin{figure*}[htb!]
\gridline{\fig{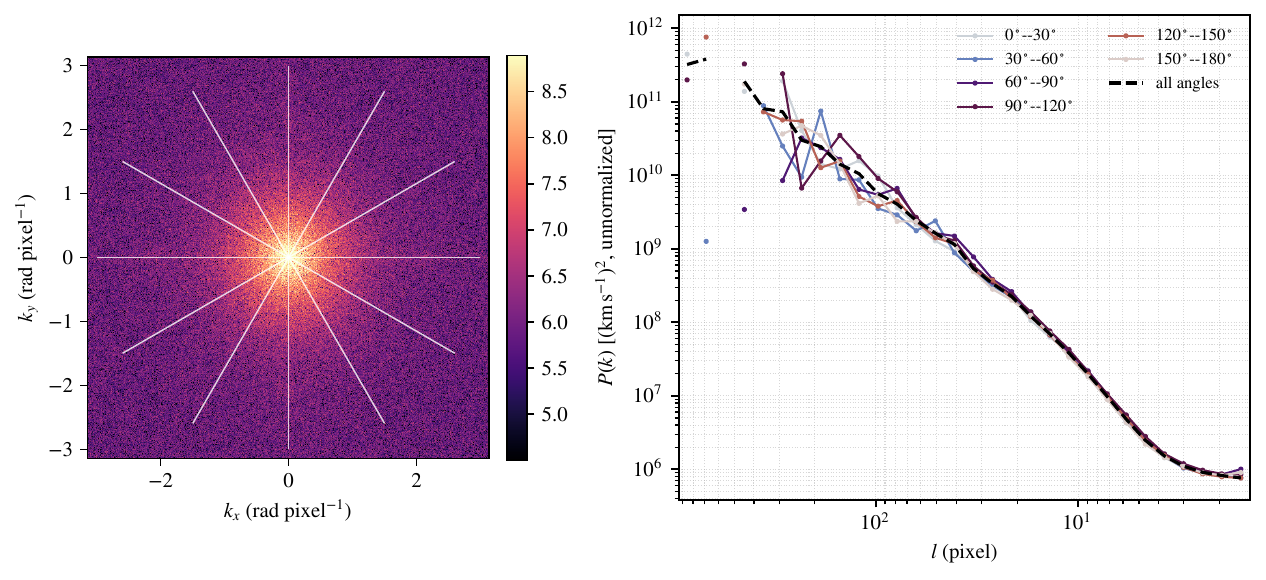}{0.7\textwidth}{(a)}}
\gridline{\fig{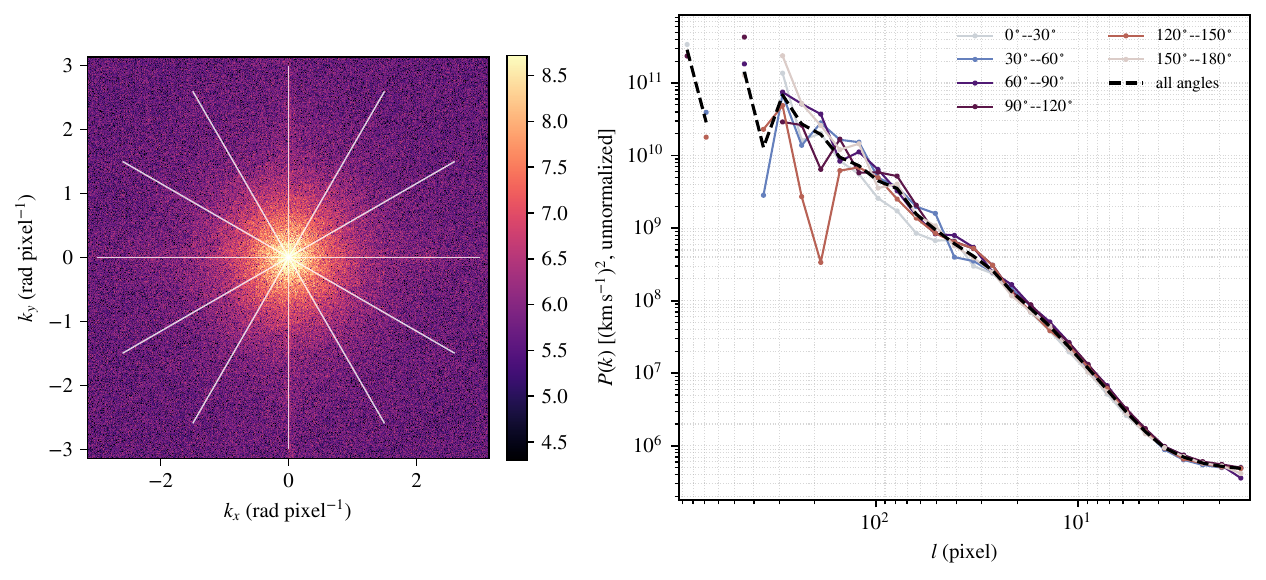}{0.7\textwidth}{(b)}}
\caption{Statistical-isotropy test for the SPSs of the two subregions shown in Figure~\ref{fig:app_isotropy_regions}. For each subregion, the two-dimensional Fourier plane is divided into six $30^{\circ}$ azimuthal sectors, and the SPS is calculated independently within each sector. The left panels show the 2D SPSs and sector boundaries, while the right panels show the corresponding sector-averaged SPSs.}
\label{fig:app_isotropy_sps}
\end{figure*}

We also test the isotropy of the SPS using the same square subregions. The two-dimensional Fourier plane is divided into six azimuthal sectors, each spanning 30\textdegree, and the SPS is calculated within each sector. The resulting 2D SPSs and the sector-averaged SPSs are shown in Figure~\ref{fig:app_isotropy_sps}. As in Figure~\ref{fig:app_isotropy_sps}, the sector-averaged SPSs closely follow the all-angle spectrum over most of the sampled spatial range. Small deviations are present only at the largest spatial scales in a few directions, where the number of independent Fourier modes is limited. Overall, the directional SPSs show no systematic azimuthal dependence in either their amplitudes or scaling behavior, indicating that the SPS statistics are approximately isotropic over the spatial scales relevant to our analysis.

\subsection{Noise-only Test} \label{app:noise_test}

We further test whether instrumental noise or the numerical analysis procedures can generate an apparent power-law scaling. Moment-0 maps are constructed from two emission-free velocity intervals, each with a width of $1~\mathrm{km\,s^{-1}}$, comparable to the typical $1$--$2~\mathrm{km\,s^{-1}}$ width of the velocity slices used in the main analysis, and analyzed using the same VSF and SPS procedures adopted for the molecular-line emission, including the same spatial sampling and Fourier-space treatment. We adopt the distance of $1~\mathrm{kpc}$ to convert angular separations to physical scales, so that the resulting VSF and SPS can be directly compared with those derived from the CO-emitting maps.

\begin{figure*}[htb!]
\centering
\includegraphics[width=0.45\textwidth]{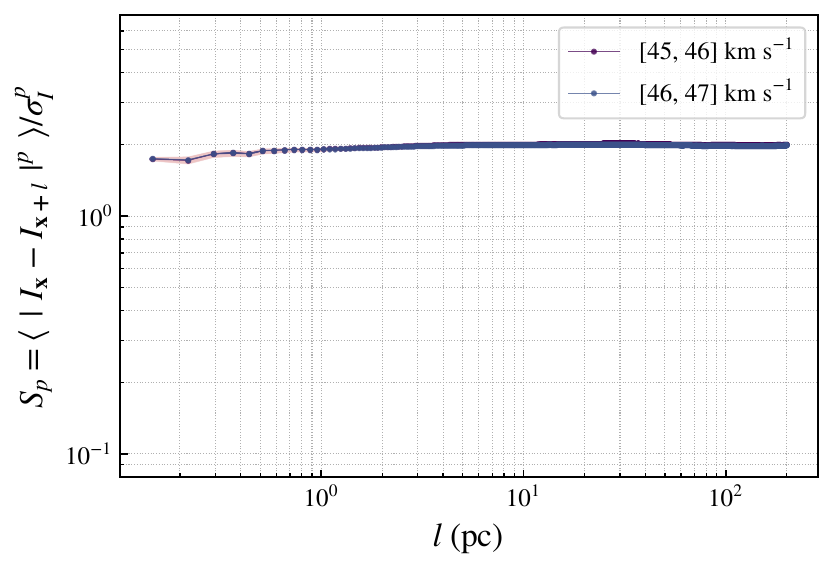}
\includegraphics[width=0.48\textwidth]{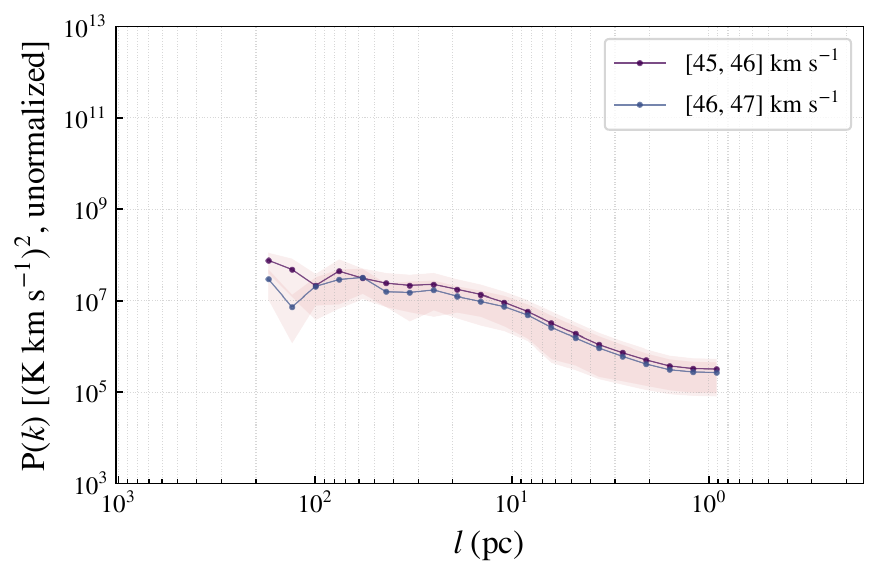}
\caption{Noise-only tests for the second-order structure function (left) and spatial power spectrum (right), calculated from a moment-0 map constructed over two emission-free velocity intervals.}
\label{fig:app_noise_test}
\end{figure*}

Figure~\ref{fig:app_noise_test} shows the resulting noise-only VSF and SPS. Neither statistic exhibits a sustained power-law interval comparable to that measured from the CO-emitting maps. The resulting noise power spectra are substantially weaker than those measured from the molecular-line emission, with the noise power being approximately one order of magnitude lower over the spatial scales relevant to our analysis. This comparison indicates that the observed VSFs and SPSs power-law behavior for the CO-emitting maps is not significantly influenced by observational noise. 


\bibliography{casecade_apj_arxiv.bbl}{}

\begin{thebibliography}{}
\expandafter\ifx\csname natexlab\endcsname\relax\def\natexlab#1{#1}\fi
\providecommand{\url}[1]{\href{#1}{#1}}
\providecommand{\dodoi}[1]{doi:~\href{http://doi.org/#1}{\nolinkurl{#1}}}
\providecommand{\doeprint}[1]{\href{http://ascl.net/#1}{\nolinkurl{http://ascl.net/#1}}}
\providecommand{\doarXiv}[1]{\href{https://arxiv.org/abs/#1}{\nolinkurl{https://arxiv.org/abs/#1}}}

\bibitem[{J. Bec \& K. Khanin(2007)Bec \& Khanin}]{bec2007burgers}
Bec, J., \& Khanin, K. 2007, \bibinfo{title}{Burgers turbulence,} Physics reports, 447, 1

\bibitem[{R. {Benzi} {et~al.}(1993){Benzi}, {Ciliberto}, {Tripiccione}, {Baudet}, {Massaioli}, \& {Succi}}]{Benzi1993}
{Benzi}, R., {Ciliberto}, S., {Tripiccione}, R., {et~al.} 1993, \bibinfo{title}{{Extended self-similarity in turbulent flows},} \pre, 48, R29, \dodoi{10.1103/PhysRevE.48.R29}

\bibitem[{S. {Boldyrev}(2002){Boldyrev}}]{Boldyrev2002a}
{Boldyrev}, S. 2002, \bibinfo{title}{{Kolmogorov-Burgers Model for Star-forming Turbulence},} \apj, 569, 841, \dodoi{10.1086/339403}

\bibitem[{J.-J. {Cai} {et~al.}(2021){Cai}, {Yang}, {Zheng}, {Yan}, {Zhang}, {Zhou}, \& {Feng}}]{Cai2021}
{Cai}, J.-J., {Yang}, J., {Zheng}, S., {et~al.} 2021, \bibinfo{title}{{Preliminary analysis on the noise characteristics of MWISP data},} Research in Astronomy and Astrophysics, 21, 304, \dodoi{10.1088/1674-4527/21/12/304}

\bibitem[{S. {Chakraborty} {et~al.}(2010){Chakraborty}, {Frisch}, \& {Ray}}]{Chakraborty2010}
{Chakraborty}, S., {Frisch}, U., \& {Ray}, S.~S. 2010, \bibinfo{title}{{Extended self-similarity works for the Burgers equation and why},} Journal of Fluid Mechanics, 649, 275, \dodoi{10.1017/S0022112010000595}

\bibitem[{A. {Chepurnov} {et~al.}(2010){Chepurnov}, {Lazarian}, {Stanimirovi{\'c}}, {Heiles}, \& {Peek}}]{Chepurnov2010}
{Chepurnov}, A., {Lazarian}, A., {Stanimirovi{\'c}}, S., {Heiles}, C., \& {Peek}, J.~E.~G. 2010, \bibinfo{title}{{Velocity Spectrum for H I at High Latitudes},} \apj, 714, 1398, \dodoi{10.1088/0004-637X/714/2/1398}

\bibitem[{M. {Chevance} {et~al.}(2023){Chevance}, {Krumholz}, {McLeod}, {Ostriker}, {Rosolowsky}, \& {Sternberg}}]{Chevance2023}
{Chevance}, M., {Krumholz}, M.~R., {McLeod}, A.~F., {et~al.} 2023, in Astronomical Society of the Pacific Conference Series, Vol. 534, Protostars and Planets VII, ed. S.~{Inutsuka}, Y.~{Aikawa}, T.~{Muto}, K.~{Tomida}, \& M.~{Tamura}, 1, \dodoi{10.48550/arXiv.2203.09570}

\bibitem[{R.~A. {Chira} {et~al.}(2019){Chira}, {Ib{\'a}{\~n}ez-Mej{\'\i}a}, {Mac Low}, \& {Henning}}]{Chira2019}
{Chira}, R.~A., {Ib{\'a}{\~n}ez-Mej{\'\i}a}, J.~C., {Mac Low}, M.~M., \& {Henning}, T. 2019, \bibinfo{title}{{How do velocity structure functions trace gas dynamics in simulated molecular clouds?},} \aap, 630, A97, \dodoi{10.1051/0004-6361/201833970}

\bibitem[{C.~L. {Dobbs} {et~al.}(2014){Dobbs}, {Krumholz}, {Ballesteros-Paredes}, {Bolatto}, {Fukui}, {Heyer}, {Mac Low}, {Ostriker}, \& {V{\'a}zquez-Semadeni}}]{Dobbs2014}
{Dobbs}, C.~L., {Krumholz}, M.~R., {Ballesteros-Paredes}, J., {et~al.} 2014, in Protostars and Planets VI, ed. H.~{Beuther}, R.~S. {Klessen}, C.~P. {Dullemond}, \& T.~{Henning}, 3--26, \dodoi{10.2458/azu_uapress_9780816531240-ch001}

\bibitem[{P. {Dutta} {et~al.}(2009){Dutta}, {Begum}, {Bharadwaj}, \& {Chengalur}}]{Dutta2009}
{Dutta}, P., {Begum}, A., {Bharadwaj}, S., \& {Chengalur}, J.~N. 2009, \bibinfo{title}{{A study of interstellar medium of dwarf galaxies using HI power spectrum analysis},} \mnras, 398, 887, \dodoi{10.1111/j.1365-2966.2009.15105.x}

\bibitem[{B.~G. {Elmegreen} \& J. {Scalo}(2004){Elmegreen} \& {Scalo}}]{ElmegreenScalo2004}
{Elmegreen}, B.~G., \& {Scalo}, J. 2004, \bibinfo{title}{{Interstellar Turbulence I: Observations and Processes},} \araa, 42, 211, \dodoi{10.1146/annurev.astro.41.011802.094859}

\bibitem[{M. Ester {et~al.}(1996)Ester, Kriegel, Sander, Xu, {et~al.}}]{Ester1996}
Ester, M., Kriegel, H.-P., Sander, J., Xu, X., {et~al.} 1996, in Kdd, Vol.~96, 226--231

\bibitem[{U. Frisch(1995)Frisch}]{Frisch1995}
Frisch, U. 1995, Turbulence: The Legacy of A. N. Kolmogorov (Cambridge University Press), \dodoi{10.1017/CBO9781139170666}

\bibitem[{K. {Fujii} {et~al.}(2021){Fujii}, {Mizuno}, {Dawson}, {Inoue}, {Torii}, {Onishi}, {Kawamura}, {Muller}, {Minamidani}, {Tsuge}, \& {Fukui}}]{Fujii2021}
{Fujii}, K., {Mizuno}, N., {Dawson}, J.~R., {et~al.} 2021, \bibinfo{title}{{Giant molecular cloud formation at the interface of colliding supershells in the large magellanic cloud},} \mnras, 505, 459, \dodoi{10.1093/mnras/stab1202}

\bibitem[{H. {Fujisaka} \& S. {Grossmann}(2001){Fujisaka} \& {Grossmann}}]{Fujisaka2001}
{Fujisaka}, H., \& {Grossmann}, S. 2001, \bibinfo{title}{{Scaling hypothesis leading to extended self-similarity in turbulence},} \pre, 63, 026305, \dodoi{10.1103/PhysRevE.63.026305}

\bibitem[{D.~A. {Green}(1993){Green}}]{Green1993}
{Green}, D.~A. 1993, \bibinfo{title}{{A power spectrum analysis of the angular scale of Galactic neutral hydrogen emission towards L = 140 deg, B = 0 deg},} \mnras, 262, 327, \dodoi{10.1093/mnras/262.2.327}

\bibitem[{J.~D. {Henshaw} {et~al.}(2020){Henshaw}, {Kruijssen}, {Longmore}, {Riener}, {Leroy}, {Rosolowsky}, {Ginsburg}, {Battersby}, {Chevance}, {Meidt}, {Glover}, {Hughes}, {Kainulainen}, {Klessen}, {Schinnerer}, {Schruba}, {Beuther}, {Bigiel}, {Blanc}, {Emsellem}, {Henning}, {Herrera}, {Koch}, {Pety}, {Ragan}, \& {Sun}}]{Henshaw2020}
{Henshaw}, J.~D., {Kruijssen}, J.~M.~D., {Longmore}, S.~N., {et~al.} 2020, \bibinfo{title}{{Ubiquitous velocity fluctuations throughout the molecular interstellar medium},} Nature Astronomy, 4, 1064, \dodoi{10.1038/s41550-020-1126-z}

\bibitem[{M.~H. {Heyer} \& C.~M. {Brunt}(2004){Heyer} \& {Brunt}}]{Heyer2004}
{Heyer}, M.~H., \& {Brunt}, C.~M. 2004, \bibinfo{title}{{The Universality of Turbulence in Galactic Molecular Clouds},} \apjl, 615, L45, \dodoi{10.1086/425978}

\bibitem[{T.~Y. {Hou} {et~al.}(1998){Hou}, {Wu}, {Chen}, \& {Zhou}}]{Hou1998}
{Hou}, T.~Y., {Wu}, X.-H., {Chen}, S., \& {Zhou}, Y. 1998, \bibinfo{title}{{Effect of finite computational domain on turbulence scaling law in both physical and spectral spaces},} \pre, 58, 5841, \dodoi{10.1103/PhysRevE.58.5841}

\bibitem[{Y. {Hu} {et~al.}(2022){Hu}, {Federrath}, {Xu}, \& {Mathew}}]{Hu2022}
{Hu}, Y., {Federrath}, C., {Xu}, S., \& {Mathew}, S.~S. 2022, \bibinfo{title}{{The velocity statistics of turbulent clouds in the presence of gravity, magnetic fields, radiation, and outflow feedback},} \mnras, 513, 2100, \dodoi{10.1093/mnras/stac972}

\bibitem[{C.-G. {Kim} {et~al.}(2006){Kim}, {Kim}, \& {Ostriker}}]{Kim2006}
{Kim}, C.-G., {Kim}, W.-T., \& {Ostriker}, E.~C. 2006, \bibinfo{title}{{Interstellar Turbulence Driving by Galactic Spiral Shocks},} \apjl, 649, L13, \dodoi{10.1086/508160}

\bibitem[{M.~I.~N. {Kobayashi} {et~al.}(2022){Kobayashi}, {Inoue}, {Tomida}, {Iwasaki}, \& {Nakatsugawa}}]{Kobayashi2022}
{Kobayashi}, M. I.~N., {Inoue}, T., {Tomida}, K., {Iwasaki}, K., \& {Nakatsugawa}, H. 2022, \bibinfo{title}{{Nature of Supersonic Turbulence and Density Distribution Function in the Multiphase Interstellar Medium},} \apj, 930, 76, \dodoi{10.3847/1538-4357/ac5a54}

\bibitem[{A. {Kolmogorov}(1941){Kolmogorov}}]{Kolmogorov1941a}
{Kolmogorov}, A. 1941, \bibinfo{title}{{The Local Structure of Turbulence in Incompressible Viscous Fluid for Very Large Reynolds' Numbers},} Akademiia Nauk SSSR Doklady, 30, 301

\bibitem[{A.~G. {Kritsuk} {et~al.}(2007){Kritsuk}, {Norman}, {Padoan}, \& {Wagner}}]{Kritsuk2007}
{Kritsuk}, A.~G., {Norman}, M.~L., {Padoan}, P., \& {Wagner}, R. 2007, \bibinfo{title}{{The Statistics of Supersonic Isothermal Turbulence},} \apj, 665, 416, \dodoi{10.1086/519443}

\bibitem[{Y. {Ma} {et~al.}(2025){Ma}, {Zhang}, {Wang}, {Fang}, {Yue}, {Chen}, {Yang}, {Du}, {Su}, {He}, {Feng}, {Sun}, {Li}, {Yan}, {Chen}, {Zhang}, \& {Zhou}}]{Ma2025}
{Ma}, Y., {Zhang}, M., {Wang}, H., {et~al.} 2025, \bibinfo{title}{{Examining Turbulence in Galactic Molecular Clouds. I. A Statistical Analysis of Velocity Structures},} \apj, 979, 65, \dodoi{10.3847/1538-4357/ad9b0e}

\bibitem[{M.-M. {Mac Low} \& R.~S. {Klessen}(2004){Mac Low} \& {Klessen}}]{MacLowKlessen2004}
{Mac Low}, M.-M., \& {Klessen}, R.~S. 2004, \bibinfo{title}{{Control of star formation by supersonic turbulence},} Reviews of Modern Physics, 76, 125, \dodoi{10.1103/RevModPhys.76.125}

\bibitem[{S.~E. {Meidt} {et~al.}(2013){Meidt}, {Schinnerer}, {Garc{\'\i}a-Burillo}, {Hughes}, {Colombo}, {Pety}, {Dobbs}, {Schuster}, {Kramer}, {Leroy}, {Dumas}, \& {Thompson}}]{Meidt2013}
{Meidt}, S.~E., {Schinnerer}, E., {Garc{\'\i}a-Burillo}, S., {et~al.} 2013, \bibinfo{title}{{Gas Kinematics on Giant Molecular Cloud Scales in M51 with PAWS: Cloud Stabilization through Dynamical Pressure},} \apj, 779, 45, \dodoi{10.1088/0004-637X/779/1/45}

\bibitem[{S.~E. {Meidt} {et~al.}(2018){Meidt}, {Leroy}, {Rosolowsky}, {Kruijssen}, {Schinnerer}, {Schruba}, {Pety}, {Blanc}, {Bigiel}, {Chevance}, {Hughes}, {Querejeta}, \& {Usero}}]{Meidt2018}
{Meidt}, S.~E., {Leroy}, A.~K., {Rosolowsky}, E., {et~al.} 2018, \bibinfo{title}{{A Model for the Onset of Self-gravitation and Star Formation in Molecular Gas Governed by Galactic Forces. I. Cloud-scale Gas Motions},} \apj, 854, 100, \dodoi{10.3847/1538-4357/aaa290}

\bibitem[{A.~K. {Mittal} {et~al.}(2023){Mittal}, {Babler}, {Stanimirovi{\'c}}, \& {Pingel}}]{Mittal2023}
{Mittal}, A.~K., {Babler}, B.~L., {Stanimirovi{\'c}}, S., \& {Pingel}, N. 2023, \bibinfo{title}{{Neutral Hydrogen (H I) 21 cm as a Probe: Investigating Spatial Variations in Interstellar Turbulent Properties},} \apj, 958, 192, \dodoi{10.3847/1538-4357/ad0464}

\bibitem[{P. {Mocz} {et~al.}(2017){Mocz}, {Burkhart}, {Hernquist}, {McKee}, \& {Springel}}]{Mocz2017}
{Mocz}, P., {Burkhart}, B., {Hernquist}, L., {McKee}, C.~F., \& {Springel}, V. 2017, \bibinfo{title}{{Moving-mesh Simulations of Star-forming Cores in Magneto-gravo-turbulence},} \apj, 838, 40, \dodoi{10.3847/1538-4357/aa6475}

\bibitem[{M. {Nandakumar} \& P. {Dutta}(2020){Nandakumar} \& {Dutta}}]{Nandakumar2020}
{Nandakumar}, M., \& {Dutta}, P. 2020, \bibinfo{title}{{Evidence of large-scale energy cascade in the spiral galaxy NGC 5236},} \mnras, 496, 1803, \dodoi{10.1093/mnras/staa1651}

\bibitem[{M. {Nandakumar} \& P. {Dutta}(2023){Nandakumar} \& {Dutta}}]{Nandakumar2023}
{Nandakumar}, M., \& {Dutta}, P. 2023, \bibinfo{title}{{Large-scale turbulence cascade in the spiral galaxy NGC 6946},} \mnras, 526, 4690, \dodoi{10.1093/mnras/stad3042}

\bibitem[{P. {Padoan} {et~al.}(2006){Padoan}, {Juvela}, {Kritsuk}, \& {Norman}}]{Padoan2006}
{Padoan}, P., {Juvela}, M., {Kritsuk}, A., \& {Norman}, M.~L. 2006, \bibinfo{title}{{The Power Spectrum of Supersonic Turbulence in Perseus},} \apjl, 653, L125, \dodoi{10.1086/510620}

\bibitem[{N.~M. {Pingel} {et~al.}(2018){Pingel}, {Lee}, {Burkhart}, \& {Stanimirovi{\'c}}}]{Pingel2018}
{Pingel}, N.~M., {Lee}, M.-Y., {Burkhart}, B., \& {Stanimirovi{\'c}}, S. 2018, \bibinfo{title}{{Multi-phase Turbulence Density Power Spectra in the Perseus Molecular Cloud},} \apj, 856, 136, \dodoi{10.3847/1538-4357/aab34b}

\bibitem[{S.~B. {Pope}(2000){Pope}}]{Pope2000}
{Pope}, S.~B. 2000, {Turbulent Flows}

\bibitem[{M.~J. {Reid} {et~al.}(2019){Reid}, {Menten}, {Brunthaler}, {Zheng}, {Dame}, {Xu}, {Li}, {Sakai}, {Wu}, {Immer}, {Zhang}, {Sanna}, {Moscadelli}, {Rygl}, {Bartkiewicz}, {Hu}, {Quiroga-Nu{\~n}ez}, \& {van Langevelde}}]{Reid2019}
{Reid}, M.~J., {Menten}, K.~M., {Brunthaler}, A., {et~al.} 2019, \bibinfo{title}{{Trigonometric Parallaxes of High-mass Star-forming Regions: Our View of the Milky Way},} \apj, 885, 131, \dodoi{10.3847/1538-4357/ab4a11}

\bibitem[{W.~W. {Roberts}(1969){Roberts}}]{Roberts1969}
{Roberts}, W.~W. 1969, \bibinfo{title}{{Large-Scale Shock Formation in Spiral Galaxies and its Implications on Star Formation},} \apj, 158, 123, \dodoi{10.1086/150177}

\bibitem[{W. {Shan} {et~al.}(2012){Shan}, {Yang}, {Shi}, {Yao}, {Zuo}, {Lin}, {Chen}, {Zhang}, {Duan}, {Cao}, {Li}, {Li}, {Liu}, \& {Zhong}}]{Shan2012}
{Shan}, W., {Yang}, J., {Shi}, S., {et~al.} 2012, \bibinfo{title}{{Development of Superconducting Spectroscopic Array Receiver: A Multibeam 2SB SIS Receiver for Millimeter-Wave Radio Astronomy},} IEEE Transactions on Terahertz Science and Technology, 2, 593, \dodoi{10.1109/TTHZ.2012.2213818}

\bibitem[{Z.-S. {She} \& E. {Leveque}(1994){She} \& {Leveque}}]{She1994}
{She}, Z.-S., \& {Leveque}, E. 1994, \bibinfo{title}{{Universal scaling laws in fully developed turbulence},} \prl, 72, 336, \dodoi{10.1103/PhysRevLett.72.336}

\bibitem[{Y. {Su} {et~al.}(2019){Su}, {Yang}, {Zhang}, {Gong}, {Wang}, {Zhou}, {Wang}, {Chen}, {Sun}, {Chen}, {Xu}, \& {Jiang}}]{Su2019}
{Su}, Y., {Yang}, J., {Zhang}, S., {et~al.} 2019, \bibinfo{title}{{The Milky Way Imaging Scroll Painting (MWISP): Project Details and Initial Results from the Galactic Longitudes of 25.{\textdegree}8-49.{\textdegree}7},} \apjs, 240, 9, \dodoi{10.3847/1538-4365/aaf1c8}

\bibitem[{K. {Sun} {et~al.}(2006){Sun}, {Kramer}, {Ossenkopf}, {Bensch}, {Stutzki}, \& {Miller}}]{Sun2006}
{Sun}, K., {Kramer}, C., {Ossenkopf}, V., {et~al.} 2006, \bibinfo{title}{{A KOSMA 7 deg$^{2}$ $^{13}$CO 2-1 and $^{12}$CO 3-2 survey of the Perseus cloud. I. Structure analysis},} \aap, 451, 539, \dodoi{10.1051/0004-6361:20054256}

\bibitem[{E. {V{\'a}zquez-Semadeni}(2025){V{\'a}zquez-Semadeni}}]{Vazquez-Semadeni2025}
{V{\'a}zquez-Semadeni}, E. 2025, {Interstellar Flow and Star Formation}, \dodoi{10.1088/2514-3433/adcb28}

\bibitem[{T.~V. {Wenger} {et~al.}(2018){Wenger}, {Balser}, {Anderson}, \& {Bania}}]{Wenger2018}
{Wenger}, T.~V., {Balser}, D.~S., {Anderson}, L.~D., \& {Bania}, T.~M. 2018, \bibinfo{title}{{Kinematic Distances: A Monte Carlo Method},} \apj, 856, 52, \dodoi{10.3847/1538-4357/aaaec8}

\bibitem[{Y. {Xu} {et~al.}(2013){Xu}, {Li}, {Reid}, {Menten}, {Zheng}, {Brunthaler}, {Moscadelli}, {Dame}, \& {Zhang}}]{Xu2013}
{Xu}, Y., {Li}, J.~J., {Reid}, M.~J., {et~al.} 2013, \bibinfo{title}{{On the Nature of the Local Spiral Arm of the Milky Way},} \apj, 769, 15, \dodoi{10.1088/0004-637X/769/1/15}

\bibitem[{Q.-Z. {Yan} {et~al.}(2021{\natexlab{a}}){Yan}, {Yang}, {Su}, {Sun}, {Xu}, {Wang}, {Zhou}, \& {Wang}}]{Yan2021a}
{Yan}, Q.-Z., {Yang}, J., {Su}, Y., {et~al.} 2021{\natexlab{a}}, \bibinfo{title}{{Improved Measurements of Molecular Cloud Distances Based on Global Search},} \apj, 922, 8, \dodoi{10.3847/1538-4357/ac214f}

\bibitem[{Q.-Z. {Yan} {et~al.}(2021{\natexlab{b}}){Yan}, {Yang}, {Sun}, {Su}, {Xu}, {Wang}, {Zhou}, \& {Wang}}]{Yan2021b}
{Yan}, Q.-Z., {Yang}, J., {Sun}, Y., {et~al.} 2021{\natexlab{b}}, \bibinfo{title}{{Distances to molecular clouds in the second Galactic quadrant},} \aap, 645, A129, \dodoi{10.1051/0004-6361/202039768}

\bibitem[{J. {Yang} {et~al.}(2026){Yang}, {Yan}, {Su}, {Zhang}, {Zhou}, {Sun}, {Ao}, {Chen}, {Chen}, {Du}, {Fang}, {Gong}, {Jiang}, {Jin}, {Ju}, {Li}, {Li}, {Liu}, {Lu}, {Luo}, {Ma}, {Mao}, {Sun}, {Wang}, {Wang}, {Wang}, {Wang (Qinghai)}, {Wang}, {Xu}, {Xu}, {Yan}, {Yan}, {Yuan}, {Zhang}, \& {Zhang}}]{Yang2026}
{Yang}, J., {Yan}, Q.-Z., {Su}, Y., {et~al.} 2026, \bibinfo{title}{{The Milky Way Imaging Scroll Painting Survey: Data Release 1},} \apjs, 282, 65, \dodoi{10.3847/1538-4365/ae29e7}

\end{thebibliography}
\bibliographystyle{aasjournalv7}



\end{document}